\documentclass[onecolumn,showpacs,showkeys,aps,prd]{revtex4}

\usepackage{latexsym}
\usepackage{amsmath,amssymb}
\usepackage{tikz}
\usepackage{color}

\newcommand{\realni}{\ensuremath{\mathbb{R}}}
\newcommand{\cD}{{\cal D}}
\newcommand{\cG}{{\cal G}}
\newcommand{\cH}{{\cal H}}
\newcommand{\cJ}{{\cal J}}

\newcommand{\cM}{{\cal M}}

\newcommand{\fg}{{\mathfrak g}}
\newcommand{\fh}{{\mathfrak h}}
\newcommand{\diag}{{\mathop{\rm diag}\nolimits}}
\newcommand{\raz}{ \rule{0ex}{2.5ex} }
\newcommand{\del}{\partial}
\newcommand{\ds}{\displaystyle}
\newcommand{\lc}{\varepsilon}
\newcommand{\hodge}{\star}
\newcommand{\pb}[2]{\{\,  {#1} \,,\, {#2} \,\}  }

\newcommand{\rmd}{d}

\newcommand{\ubar}[1]{\underline{#1}} % underbar

\newcommand{\te}[2]{ \tilde{e}{}^{#1}{}_{#2}\, }
\newcommand{\teu}[2]{ \tilde{e}{}^{#1 #2}\, }

\newcommand{\tg}{ \tilde{g}{} }

\newcommand{\ed}[1]{ e_{#1} }
\newcommand{\eu}[1]{ e^{#1} }

\newcommand{\tedet}{ e_3 }
\newcommand{\gedet}{ g_3 }

\newcommand{\orto}{\bot}
\newcommand{\para}{\parallel}

\begin{document}

\title{Hamiltonian analysis of the $BFCG$ formulation of General Relativity}

\author{Aleksandar Mikovi\'c}
\email{amikovic@ulusofona.pt}
\affiliation{Departamento de Matem\'atica Universidade Lus\'ofona de Humanidades e Tecnologias, Av. do Campo Grande 376, 1749-024 Lisboa, Portugal}
\affiliation{Grupo de F\'isica Matem\'atica, Faculdade de Ci\^ encias da Universidade de Lisboa, Campo Grande, Edif\'icio C6, 1749-016  Lisboa, Portugal}

\author{Miguel A. Oliveira}
\email{masm.oliveira@gmail.com}
\affiliation{Grupo de F\'isica Matem\'atica, Faculdade de Ci\^ encias da Universidade de Lisboa, Campo Grande, Edif\'icio C6, 1749-016  Lisboa, Portugal}

\author{Marko Vojinovi\'c}
\email{vmarko@ipb.ac.rs}
\affiliation{Group for Gravitation, Particles and Fields, Institute of Physics, University of Belgrade, Pregrevica 118, 11080 Belgrade, Serbia}



\begin{abstract}
We perform the complete Hamiltonian analysis of the BFCG action for General Relativity. We determine all the constraints of the theory and classify them into the first-class and the second-class constraints. We also show how the canonical formulation of BFCG General Relativity reduces to the Einstein-Cartan and triad canonical formulations. The reduced phase space analysis also gives a 2-connection which is suitable for the construction of a spin-foam basis which will be a categorical generalization of the spin-network basis from Loop Quantum Gravity.
\end{abstract}

\maketitle

\section{\label{SecIntroduction}Introduction}

Among the fundamental problems of modern theoretical physics, by far the most prominent one is the construction of the tentative theory of quantum gravity (QG). There are many approaches to QG, one of which is called Loop Quantum Gravity (LQG), see \cite{lqg,T,sf,RV}. As with any other physical system, the quantization of the gravitational field can be performed either canonically, using the Hamiltonian framework, or covariantly, using the Lagrangian, i.e., the path integral framework. Within the LQG approach, in the canonical framework \cite{T} one chooses the connection variables and their momenta as fundamental fields for gravity, and uses them to construct an appropriate physical Hilbert space, giving rise to the spin-network states. In the covariant framework, one puts the connection variables onto a spacetime triangulation, see \cite{sf,RV}, and uses this construction to define a path integral for gravity, giving rise to the spin-foam (SF) models.

%The problem of quantization of gravitational field is arguably the most important fundamental problem in modern theoretical physics. One of the approaches to the construction of a theory of quantum gravity is called Loop Quantum Gravity (LQG), see \cite{lqg,T,sf,RV}. Research in LQG is focused on two complementary fronts --- canonical and covariant. In the canonical approach \cite{T}, one rewrites General Relativity (GR) in the Hamiltonian form, by using connection variables and their momenta, with the goal of constructing the appropriate physical Hilbert space. This amounts to working with the the spin-network states. In the covariant approach, the aim is to perform the path-integral quantization of GR by using the connection variables, and this requires using a spacetime triangulation where the triangles carry spin labels, see \cite{sf,RV}. The resulting theories are called spin-foam (SF) models. 

The $BFCG$ formulation of GR \cite{MV} was invented in order to find a categorical generalization of the SF models. A categorical generalization of a SF model is called a spin-cube model, since the path integral is based on a colored 3-complex where the colors are the representations of a $2$-group \cite{MV,M}. The 2-group, see \cite{BaezHuerta2011} for a review and references, replaces the Lorentz group, and becomes the fundamental algebraic structure. The reason for introducing spin-cube models was that the SF models have two problems. One problem is that the classical limit of a SF model is described by the area-Regge action \cite{mvea,M}. The second problem is that the fermions cannot be coupled to a SF model \cite{MV}. These two problems are caused by the fact that the tetrads are absent from the Plebanski action, see \cite{Plebanski,DePietriFreidel,LivineOriti,sf}, which is used as the classical action to build the SF amplitudes. The $BFCG$ action for GR is a categorical generalization of the Plebanski action, and the $BFCG$ action contains both the $B$ field and the tetrads \cite{MV}.

The path integral quantization of $BFCG$ GR reduces to the Regge path integral \cite{M}. However, in the case of the canonical quantization, it is not known what kind of theories can be obtained. It was argued in \cite{MO} that a spin-foam basis should exist, as a categorical generalization of the spin-network basis from LQG, but in order to rigorously prove such a statement, one needs a canonical formulation of the $BFCG$ GR theory. The canonical analysis of $BFCG$ GR action is much more complicated than the canonical analysis of the Einstein-Hilbert action. One can see what kind of canonical analysis will be necessary from the canonical analysis of simpler but related actions given by the unconstrained $BFCG$ action \cite{MOV} or the Einstein-Cartan action \cite{B}.

In this paper we present the Hamiltonian analysis of the $BFCG$ GR theory in full detail. Despite being straightforward, the calculations involved are quite nontrivial, so it is important to perform the full analysis in a systematic manner. Due to the amount of material presented, subsequent topics such as quantization schemes and similar have been postponed for future work, while the present paper deals only with the canonical structure of the classical theory.

The paper is organized as follows. In section \ref{SecII} we give an overview of the $BFCG$ GR action, discuss the Lagrange equations of motion, and prepare for the Hamiltonian analysis. The first part of the Hamiltonian analysis is done in section \ref{SecIII}. We evaluate the conjugate momenta for the fields, obtain the primary constraints and construct the Hamiltonian of the theory. Then we impose consistency conditions on on all constraints in turn, giving rise to a full set of primary, secondary and tertiary constraints, along with some determined Lagrange multipliers. Section \ref{SecIV} is devoted to the second part of the Hamiltonian analysis --- the separation of the constraints into first and second class, computing their algebra, and determining the number of physical degrees of freedom. Building on these results, in section \ref{SecV} we discuss various avenues for the elimination of the second class constraints from the theory, gauge fixing conditions and the analysis of the first class constraints, and the resulting possible reductions of the phase space of the theory. Section \ref{SecVI} contains our concluding remarks, discussion of the results and future lines of research. The Appendix contains four sections with a lot of technical details about the calculations performed in the main text.

Our notation and conventions are as follows. The spacetime indices are denoted with lowercase Greek alphabet letters from the middle of the alphabet $\lambda, \mu,\nu,\rho,\dots$ and take the values $0,1,2,3$. When discussing the foliation of spacetime into space and time, the spacetime indices are split as $\mu = (0,i)$, where the lowercase indices from the middle of the Latin alphabet $i,j,k,\dots$ take only spacelike values $1,2,3$. The Poincar\'e group indices are denoted with lowercase letters from the beginning of the Latin alphabet, $a,b,c,\dots$ and take the values $0,1,2,3$, while their spacelike counterparts are denoted by the lower-case Greek letters from the beginning of the alphabet $\alpha,\beta,\dots$, and take the values $1,2,3$. The group indices are raised and lowered with the Minkowski metric $\eta_{ab} = \diag (-1,1,1,1)$. Capital Latin indices $A,B,C,\dots$ represent multi-index notation, and are used to count the second class constraints, fields and momenta, and various other objects, depending on the context. Antisymmetrization is denoted with the square brackets around the indices with the $1/2$ factor, $X_{[ab]} \equiv \left( X_{ab} - X_{ba} \right) /2$. In order to simplify the notation involving Poisson brackets, we will adopt the following convention. The left quantity in every Poisson bracket is assumed to be evaluated at the point $x=(t,\vec{x})$, while the right quantity at the point $y=(t,\vec{y})$. In addition, we use the shorthand notation for the 3-dimensional Dirac delta function $\delta^{(3)}\equiv \delta^{(3)}(\vec{x}-\vec{y})$. For example, an expression
\begin{equation}
\pb{U^\alpha(t,\vec{x})}{V^\beta(t,\vec{y})} = W^{\alpha\beta}(t,\vec{x}) \delta^{(3)}(\vec{x}-\vec{y}) + Z^{\alpha\beta i}(t,\vec{x}) \,\partial_i \delta^{(3)}(\vec{x}-\vec{y}) \,,
\end{equation}
where $\partial_i = \partial/\partial x^i$, can be written more compactly as
\begin{equation}
\pb{U^\alpha}{V^\beta} = W^{\alpha\beta} \delta^{(3)} + Z^{\alpha\beta i} \,\partial_i \delta^{(3)}\,,
\end{equation}
usually without any ambiguity. In the rare ambiguous cases, the expressions will be written more explicitly. This notation will be used systematically unless stated otherwise.

\section{\label{SecII}$BFCG$ action for GR}

Given a Lie group $\cG$ and its Lie algebra $\fg$, and the $\fg$-valued connection one-form $A$ on a spacetime manifold $\cM$, the $BF$ action (see \cite{BFreview} for a review and applications to gravity)
\begin{equation}
S_{BF} = \int_{\cM} \langle B \wedge F \rangle_{\fg}\,,
\end{equation}
describes the dynamics of flat connections, where $F=dA+A\wedge A$ is the curvature two-form. $B$ is a $\fg$-valued Lagrange multiplier two-form and $\langle\;\,,\;\rangle_{\fg}$ represents the invariant nondegenerate symmetric bilinear form in $\fg$. The $BF$ theory relevant for the construction of spin-foam models is based on the Lorentz group $SO(3,1)$. A categorical generalization of the $BF$ theory is based on the concept of a strict $2$-group, which is a pair of groups $(\cG ,\cH)$ with certain maps between them (see \cite{BaezHuerta2011} for details). The corresponding theory of flat 2-connections is called the $BFCG$ theory \cite{GirelliPfeifferPopescu2008,MM}, and its dynamics is given by the action
\begin{equation} \label{PureBFCGaction}
S_{BFCG} = \int_{\cM} \left[\langle B \wedge F \rangle_{\fg} + \langle C \wedge G \rangle_{\fh}\right]\,.
\end{equation}
The second term in (\ref{PureBFCGaction}) consists of a $\fh$-valued one-form Lagrange multiplier $C$, and a curvature three-form $G = d\beta + A\wedge \beta$ for the $\fh$-valued two-form $\beta$, where $\fh$ is the Lie algebra of the group $\cH$. The pair $(A,\beta)$ is called the $2$-connection for the $2$-group, while the pair $(F,G)$ is the corresponding $2$-curvature. The $\langle\;\,,\;\rangle_{\fh}$ is the invariant nondegenerate symmetric bilinear form in $\fh$, which is $\fg$-invariant.

The Poincar\'e $2$-group, defined by $\cG = SO(3,1)$ and $\cH=\realni^4$, is relevant for GR since the Einstein equations can be obtained from a constrained $BFCG$ action \cite{MV}, given by
\begin{equation} \label{SpincubeDejstvo}
S_{GR} = \int_{\cM} \left[\langle B \wedge R \rangle_{\fg} + \langle e \wedge G \rangle_{\fh} - \langle \phi \wedge \left(B - \hodge (e\wedge e) \right) \rangle_{\fg}\right] \, .
\end{equation}
Here we have relabeled $C\equiv e$ and $F\equiv R$, since in the case of the Poincar\'e $2$-group these fields have the interpretation of the tetrad field and the curvature two-form for the spin connection $A\equiv \omega$. The $\fg$-valued two-form $\phi$ is an additional Lagrange multiplier, featuring in the simplicity constraint term. The $\hodge$ is the Hodge dual operator for the Minkowski space.

The action (\ref{SpincubeDejstvo}) can be written as
\begin{equation} \label{BFCGdejstvo}
S_{GR} = \int_{\cM} \left[B_{ab} \wedge R^{ab} + e^a \wedge G_a - \phi^{ab} \wedge \left( B_{ab} - \lc_{abcd}\, e^c\wedge e^d \right)\right] \,,
\end{equation}
where the curvatures $R^{ab}$ and $G^a$ are given by
\begin{equation} \label{DefinicijaKrivine}
R^{ab} = d\omega^{ab} + \omega^a{}_{c} \wedge \omega^{cb}\,,
\end{equation}
\begin{equation} \label{DefinicijaGkrivine}
G^a = \nabla \beta^a \equiv d\beta^a + \omega^a{}_b \wedge \beta^b\,.
\end{equation}
The action (\ref{BFCGdejstvo}) can even be extended to include the cosmological constant, and it is related to the MacDowell-Mansouri action \cite{MacDowellMansouri,Smolin,LingSmolin,SmolinStarodubtsev,Wise}, see Appendix \ref{AppE} for details.

It is convenient to introduce the torsion 2-form
\begin{equation} \label{DefinicijaTorzije}
T^a = \nabla e^a \equiv de^a + \omega^a{}_b \wedge e^b\,,
\end{equation}
so that one can rewrite the action as
\begin{equation} \label{PGTdejstvo}
S_{PGT} = \int_{\cM} \left[ B_{ab} \wedge R^{ab} + \beta^a \wedge T_a - \phi^{ab} \wedge \left( B_{ab} - \lc_{abcd}\, e^c\wedge e^d \right) \right] \,.
\end{equation}
by using the integration by parts. The action (\ref{PGTdejstvo}) is a constrained $BF$ action for the Poincar\'e group, since the tetrads and the spin connection can be considered as components of a Poincar\'e group connection, while the curvature and the torsion are the components of the Poincar\'e group curvature \cite{MO}. This equivalence of a Poincar\'e gauge theory formulation to a 2-group gauge theory formulation is specific to $4$ spacetime dimensions only.

The relationship between the topological, unconstrained versions of the actions (\ref{BFCGdejstvo}) and (\ref{PGTdejstvo}) has been discussed in detail in \cite{MOV}. There, a real parameter $\xi$ was introduced to interpolate between the two actions, the full Hamiltonian analysis was performed, and the implications of the parameter $\xi$ for the structure of the resulting phase space were studied in detail. It is noteworthy that the actions (\ref{BFCGdejstvo}) and (\ref{PGTdejstvo}) differ from the actions discussed in \cite{MOV} only by the presence of the simplicity constraint term, which is the same for both actions and does not contain any time derivatives. Therefore, the presence of the simplicity constraint does not change any results of \cite{MOV} pertaining to the $\xi$ parameter, and all conclusions related to $\xi$ given in \cite{MOV} carry over unmodified to the constrained actions (\ref{BFCGdejstvo}) and (\ref{PGTdejstvo}) discussed in this paper. Given this situation, we opt not to introduce and discuss the $\xi$ parameter again in this paper, and refer the reader to \cite{MOV} instead.

It is clear that the actions (\ref{BFCGdejstvo}) and (\ref{PGTdejstvo}) give rise to the same set of equations of motion, since these do not depend on the boundary. Taking the variation of (\ref{BFCGdejstvo}) with respect to all the variables, one obtains
\begin{eqnarray}
\delta B:      & R^{ab} - \phi^{ab} = 0\,, \label{JKzaR} \\
\delta \beta:  & T^a = 0 \,, \label{JKzaT} \\
\delta e:      & G_a + 2\lc_{abcd}\, \phi^{bc} \wedge e^d = 0 \,, \label{JKzaG} \\
\delta \omega: & \nabla B^{ab} - e^{[a}\wedge \beta^{b]} = 0 \,, \label{JKzaNablaB} \\
\delta \phi:   & B_{ab} - \lc_{abcd}\, e^c\wedge e^d = 0 \,, \label{JKzaB} \\
\nonumber
\end{eqnarray}
where the covariant exterior derivative of $B^{ab}$ is defined as
\begin{equation}
\nabla B^{ab} \equiv d B^{ab} + \omega^a{}_c \wedge B^{cb} + \omega^b{}_c \wedge B^{ac}\,.
\end{equation}
One can simplify the equations of motion in the following way. Taking the covariant exterior derivative of (\ref{JKzaB}) and using (\ref{JKzaT}) one obtains $\nabla B^{ab}=0$. Substituting this into (\ref{JKzaNablaB}) one further obtains $e^{[a} \wedge \beta^{b]} = 0$. Under the assumption that $\det (e^a{}_{\mu}) \neq 0$, it follows that $\beta^a = 0$ (see Appendix in \cite{MV} for proof), and therefore also $G^a=0$. As a consequence, we see that the equations of motion (\ref{JKzaR}) -- (\ref{JKzaB}) are equivalent to the following system:
\begin{itemize}
\item the equation that determines the multiplier $\phi^{ab}$ in terms of curvature,
\begin{equation} \label{JKzaPhiBolja}
\phi^{ab} = R^{ab} \,,
\end{equation}
\item the equation that determines the multiplier $B^{ab}$ in terms of tetrads,
\begin{equation} \label{JKzaBbolja}
B_{ab} = \lc_{abcd}\, e^c\wedge e^d \,, 
\end{equation}
\item the equation that determines $\beta^a$,
\begin{equation} \label{JKzaBetaBolja}
\beta^a = 0 \,,
\end{equation}
\item the equation for the torsion,
\begin{equation} \label{JKzaTbolja}
T^a = 0 \,,
\end{equation}
\item and the Einstein field equation,
\begin{equation} \label{JKzaEiRbolja}
\lc_{abcd} \, R^{bc} \wedge e^d = 0 \,.
\end{equation}
\end{itemize}

Finally, for the convenience of the Hamiltonian analysis, we need to rewrite both the action and the equations of motion in a local coordinate frame. Choosing $dx^{\mu}$ as basis one-forms, we can expand the fields in the standard fashion:
\begin{equation} \label{KomponenteTetradeIkoneksije}
e^a = e^a{}_{\mu}dx^{\mu}\,, \qquad
\omega^{ab} = \omega^{ab}{}_{\mu} dx^{\mu}\,,
\end{equation}
\begin{equation} \label{KomponenteBiBeta}
B^{ab} = \frac{1}{2} B^{ab}{}_{\mu\nu} dx^{\mu} \wedge dx^{\nu}\,, \qquad
\beta^a = \frac{1}{2}\beta^a{}_{\mu\nu} dx^{\mu} \wedge dx^{\nu}\,, \qquad
\phi^{ab} = \frac{1}{2}\phi^{ab}{}_{\mu\nu} dx^{\mu} \wedge dx^{\nu}\,.
\end{equation}
Similarly, the field strengths for $\omega$, $e$ and $\beta$ are
\begin{equation} \label{KomponenteSvihKrivina}
\begin{array}{ccl}
R^{ab} & = & \ds \frac{1}{2} R^{ab}{}_{\mu\nu} dx^{\mu} \wedge dx^{\nu}\,, \vphantom{\ds\int}\\
T^a & = & \ds \frac{1}{2} T^a{}_{\mu\nu} dx^{\mu} \wedge dx^{\nu}\,, \vphantom{\ds\int}\\
G^a & = & \ds \frac{1}{6} G^a{}_{\mu\nu\rho} dx^{\mu} \wedge dx^{\nu} \wedge dx^{\rho}\,. \vphantom{\ds\int}\\
\end{array}
\end{equation}
Using the relations (\ref{DefinicijaKrivine}), (\ref{DefinicijaGkrivine}) and  (\ref{DefinicijaTorzije}), we can write the component equations
\begin{equation} \label{KomponenteJacinaPolja}
\begin{array}{ccl}
R^{ab}{}_{\mu\nu} & = & \del_{\mu} \omega^{ab}{}_{\nu} - \del_{\nu} \omega^{ab}{}_{\mu} + \omega^a{}_{c\mu} \omega^{cb}{}_{\nu} - \omega^a{}_{c\nu} \omega^{cb}{}_{\mu}\,, \\
T^a{}_{\mu\nu} & = & \del_{\mu} e^a{}_{\nu} - \del_{\nu} e^a{}_{\mu} + \omega^a{}_{b\mu} e^b{}_{\nu} - \omega^a{}_{b\nu} e^b{}_{\mu}\,, \vphantom{\ds\int} \\
G^a{}_{\mu\nu\rho} & = & \del_{\mu} \beta^a{}_{\nu\rho} + \del_{\nu} \beta^a{}_{\rho\mu} + \del_{\rho} \beta^a{}_{\mu\nu}
+ \omega^a{}_{b\mu} \beta^b{}_{\nu\rho} + \omega^a{}_{b\nu} \beta^b{}_{\rho\mu} + \omega^a{}_{b\rho} \beta^b{}_{\mu\nu}\,. \\
\end{array}
\end{equation}

Substituting expansions (\ref{KomponenteTetradeIkoneksije}), (\ref{KomponenteBiBeta}) and (\ref{KomponenteSvihKrivina}) into the action, we obtain
\begin{equation} \label{DejstvoUkomponentama}
S = \int_{\cM} \rmd^4x\, \lc^{\mu\nu\rho\sigma} \left[ \frac{1}{4} B_{ab\mu\nu} R^{ab}{}_{\rho\sigma} +
\frac{1}{6} e_{a\mu} G^a{}_{\nu\rho\sigma} -\frac{1}{4} \phi^{ab}{}_{\mu\nu} \left( B_{ab\rho\sigma} - 2\lc_{abcd}\, e^c{}_{\rho} e^d{}_{\sigma} \right)
 \right]\,.
\end{equation}
Assuming that the spacetime manifold has the topology $\cM = \Sigma\times\realni$, where $\Sigma$ is a 3-dimensional spacelike hypersurface, from the above action we can read off the Lagrangian, which is the integral of the Lagrangian density over the hypersurface $\Sigma$:
\begin{equation} \label{Lagranzijan}
L = \int_{\Sigma} \rmd^3x\, \lc^{\mu\nu\rho\sigma} \left[ \frac{1}{4} B_{ab\mu\nu} R^{ab}{}_{\rho\sigma} +
\frac{1}{6} e_{a\mu} G^a{}_{\nu\rho\sigma} -\frac{1}{4} \phi^{ab}{}_{\mu\nu} \left( B_{ab\rho\sigma} - 2\lc_{abcd} \, e^c{}_{\rho} e^d{}_{\sigma} \right)
 \right]\,.
\end{equation}
Finally, the component form of equations of motion (\ref{JKzaPhiBolja}) -- (\ref{JKzaEiRbolja}) is:
\begin{equation}
\begin{array}{c}
\ds \phi^{ab}{}_{\mu\nu} = R^{ab}{}_{\mu\nu} \,, \qquad B_{ab\mu\nu} = 2 \lc_{abcd}\, e^c{}_{\mu} e^d{}_{\nu}\,, \qquad \beta^a{}_{\mu\nu} = 0\,, \vphantom{\ds\int} \\
\ds T^a{}_{\mu\nu} = 0\,, \qquad 
 \lc^{\lambda\mu\nu\rho} \lc_{abcd}\, R^{bc}{}_{\mu\nu} e^d{}_{\rho} = 0\,. \vphantom{\ds\int} \\
\end{array}
\end{equation}

\section{\label{SecIII}Hamiltonian analysis}

Now we turn to the Hamiltonian analysis. A detailed review of the general formalism can be found in \cite{B}, Chapter V. In addition, a good pedagogical example of the Hamiltonian analysis which is relevant for our case is the topological $BFCG$ gravity \cite{MOV}.

\subsection{Primary constraints and the Hamiltonian}

As a first step, we calculate the momenta $\pi$ corresponding to the field variables $B^{ab}{}_{\mu\nu}$, $\phi^{ab}{}_{\mu\nu}$, $e^a{}_{\mu}$, $\omega^{ab}{}_{\mu}$ and $\beta^a{}_{\mu\nu}$. Differentiating the action (\ref{DejstvoUkomponentama}) with respect to the time derivative of the appropriate fields, we obtain the momenta as follows:
\begin{equation}
\begin{array}{lclcl}
\pi(B)_{ab}{}^{\mu\nu} & \raz = & \ds \frac{\delta S}{\delta \del_0 B^{ab}{}_{\mu\nu}} & = & 0\,, \\
\pi(\phi)_{ab}{}^{\mu\nu} & \raz = & \ds \frac{\delta S}{\delta \del_0 \phi^{ab}{}_{\mu\nu}} & = & 0\,, \\
\pi(e)_{a}{}^{\mu} & \raz = & \ds \frac{\delta S}{\delta \del_0 e^a{}_{\mu}} & = & 0 \,, \vphantom{\ds\int^A} \\
\pi(\omega)_{ab}{}^{\mu} & \raz = & \ds \frac{\delta S}{\delta \del_0 \omega^{ab}{}_{\mu}} & = & \ds \lc^{0\mu\nu\rho} B_{ab\nu\rho}\,, \vphantom{\ds\int^A} \\
\pi(\beta)_a{}^{\mu\nu} & \raz = &  \ds \frac{\delta S}{\delta \del_0 \beta^a{}_{\mu\nu}} & = & - \lc^{0\mu\nu\rho} e_{a\rho}\,. \vphantom{\ds\int^A} \\
\end{array}
\end{equation}
None of the momenta can be solved for the corresponding ``velocities'', so they all give rise to primary constraints:
\begin{equation}
\begin{array}{lcl}
P(B)_{ab}{}^{\mu\nu} & \raz\equiv & \pi(B)_{ab}{}^{\mu\nu} \approx 0\,, \vphantom{\ds\frac{1}{2}} \\
P(\phi)_{ab}{}^{\mu\nu} & \raz\equiv & \pi(\phi)_{ab}{}^{\mu\nu} \approx 0\,, \vphantom{\ds\frac{1}{2}} \\
P(e)_a{}^{\mu} & \raz\equiv & \ds \pi(e)_a{}^{\mu} \approx 0\,, \vphantom{\ds\frac{1}{2}} \\
P(\omega)_{ab}{}^{\mu} & \raz\equiv & \pi(\omega)_{ab}{}^{\mu} - \lc^{0\mu\nu\rho} B_{ab\nu\rho} \approx 0\,, \vphantom{\ds\frac{1}{2}} \\
P(\beta)_a{}^{\mu\nu} & \raz\equiv & \pi(\beta)_a{}^{\mu\nu} + \lc^{0\mu\nu\rho} e_{a\rho} \approx 0\,. \vphantom{\ds\frac{1}{2}} \\
\end{array}
\end{equation}
The weak, on-shell equality is denoted ``$\approx$'', as opposed to the strong, off-shell equality which is denoted by the usual symbol ``$=$''.

Next we introduce the fundamental simultaneous Poisson brackets between the fields and their conjugate momenta,
\begin{equation}
\begin{array}{lcl}
\pb{B^{ab}{}_{\mu\nu}}{\pi(B)_{cd}{}^{\rho\sigma}} & \raz = & 4 \delta^a_{[c} \delta^b_{d]} \delta^{\rho}_{[\mu} \delta^{\sigma}_{\nu]} \delta^{(3)}\,, \\
\pb{\phi^{ab}{}_{\mu\nu}}{\pi(\phi)_{cd}{}^{\rho\sigma}} & \raz = & 4 \delta^a_{[c} \delta^b_{d]} \delta^{\rho}_{[\mu} \delta^{\sigma}_{\nu]} \delta^{(3)}\,, \\
\pb{e^a{}_{\mu}}{\pi(e)_b{}^{\nu}} & \raz = & \delta^a_b \delta^{\nu}_{\mu} \delta^{(3)}\,, \\
\pb{\omega^{ab}{}_{\mu}}{\pi(\omega)_{cd}{}^{\nu}} & \raz = & 2 \delta^a_{[c} \delta^b_{d]} \delta^{\nu}_{\mu} \delta^{(3)}\,, \\
\pb{\beta^a{}_{\mu\nu}}{\pi(\beta)_b{}^{\rho\sigma}} & \raz = & 2 \delta^a_b \delta^{\rho}_{[\mu} \delta^{\sigma}_{\nu]} \delta^{(3)}\,, \\
\end{array}
\end{equation}
and we employ them to calculate the algebra of primary constraints,
\begin{equation} \label{AlgebraPrimarnihVeza}
\begin{array}{lcl}
\pb{P(B)^{abjk}}{P(\omega)_{cd}{}^i} & = & 4 \lc^{0ijk} \delta^a_{[c} \delta^b_{d]} \delta^{(3)}, \\
\pb{P(e)^{ak}}{P(\beta)_b{}^{ij}} & = & - \lc^{0ijk} \delta^a_b \delta^{(3)}, \\
\end{array}
\end{equation}
while all other Poisson brackets vanish.

Next we construct the canonical, on-shell Hamiltonian:
\begin{equation}
H_c = \ds \int_{\Sigma} \rmd^3\vec{x} \left[ \frac{1}{4} \pi(B)_{ab}{}^{\mu\nu} \del_0 B^{ab}{}_{\mu\nu} + \frac{1}{4} \pi(\phi)_{ab}{}^{\mu\nu} \del_0 \phi^{ab}{}_{\mu\nu} + \pi(e)_a{}^{\mu} \del_0 e^a{}_{\mu}
+ \frac{1}{2} \pi(\omega)_{ab}{}^{\mu} \del_0 \omega^{ab}{}_{\mu} + \frac{1}{2} \pi(\beta)_a{}^{\mu\nu} \del_0 \beta^a{}_{\mu\nu} \right] -L \,.
\end{equation}
The factors $1/4$ and $1/2$ are introduced to prevent overcounting of variables. Using (\ref{KomponenteJacinaPolja}) and  (\ref{Lagranzijan}), one can re\-arrange the expressions such that all velocities are multiplied by primary constraints, and therefore vanish from the Hamiltonian. After some algebra, the resulting expression can be written as
\begin{equation}
\begin{array}{ccl}
H_c & = & \ds - \int_{\Sigma} \rmd^3\vec{x}\, \lc^{0ijk} \left[ \frac{1}{2} B_{ab0i} \left( R^{ab}{}_{jk}-\phi^{ab}{}_{jk} \right) +  e^a{}_0 \left( \frac{1}{6} G_{aijk} +\lc_{abcd} \, \phi^{bc}{}_{ij} e^d{}_k \right) + \right. \\
 & & \hphantom{mmmmmmm} \ds \left. + \frac{1}{2} \beta_{a0k} T^a{}_{ij} + \frac{1}{2} \omega_{ab0} \left( \nabla_i B^{ab}{}_{jk} - e^a{}_i \beta^b{}_{jk} \right) -\frac{1}{2} \phi^{ab}{}_{0i} \left( B_{abjk} - 2\lc_{abcd}\, e^c{}_j e^d{}_k \right) \right] \,, \\
\end{array}
\end{equation}
up to a boundary term. The canonical Hamiltonian does not depend on any momenta, but only on fields and their spatial derivatives. Finally, introducing Lagrange multipliers $\lambda$ for each of the primary constraints, we construct the total, off-shell Hamiltonian:
\begin{equation} \label{TotalniHamiltonijan}
\begin{array}{ccl}
H_T & = & \ds H_c + \int_{\Sigma} \rmd^3\vec{x} \left[  \frac{1}{4} \lambda(B)^{ab}{}_{\mu\nu} P(B)_{ab}{}^{\mu\nu} + \frac{1}{4} \lambda(\phi)^{ab}{}_{\mu\nu} P(\phi)_{ab}{}^{\mu\nu} + \right. \\
 & & \hphantom{mmmmmmm} \ds \left. \lambda(e)^a{}_{\mu} P(e)_a{}^{\mu} + \frac{1}{2} \lambda(\omega)^{ab}{}_{\mu} P(\omega)_{ab}{}^{\mu} + \frac{1}{2} \lambda(\beta)^a{}_{\mu\nu} P(\beta)_a{}^{\mu\nu} \right] \,. \\
\end{array}
\end{equation}

\subsection{Consistency procedure}

We proceed with the calculation of the consistency requirements for the constraints. The consistency requirement is that the time derivative of each constraint (or equivalently its Poisson bracket with the total Hamiltonian (\ref{TotalniHamiltonijan})) must vanish on-shell. This requirement can either give rise to a new constraint, or determine some multiplier, or be satisfied identically. In our case, the consistency requirements give rise to a complicated chain structure, depicted in the following diagram:
\begin{center}
\begin{tikzpicture}[yscale=0.6]
\filldraw[black] (0,10) circle (0pt) node[anchor=west] {$P(\beta)_a{}^{0i}$};
\draw[->] (2,10) -- (3,10);
\filldraw[black] (2.5,10) circle (0pt) node[anchor=south] {\footnotesize 1};
\filldraw[black] (3.5,10) circle (0pt) node[anchor=west] {$S(T)^{ai}$};
\draw[->] (5.5,10) -- (6.5,10);
\filldraw[black] (6,10) circle (0pt) node[anchor=south] {\footnotesize 15};
\filldraw[black] (7,10) circle (0pt) node[anchor=west] {$T(eR\phi)^{ai}$};

\filldraw[black] (0,9) circle (0pt) node[anchor=west] {$P(e)_a{}^i$};
\draw[->] (2,9) -- (3,9);
\filldraw[black] (2.5,9) circle (0pt) node[anchor=south] {\footnotesize 11};
\filldraw[black] (3.5,9) circle (0pt) node[anchor=west] {$S(eR\phi)^{ai}$};
\draw (5.5,9) -- (9.5,9);
\draw (9,10) -- (9.5,10);
\draw (9.5,9) -- (9.5,10);
\draw[->] (9.5,9.5) -- (10.3,9.5);
\filldraw[black] (9.9,9.5) circle (0pt) node[anchor=south] {\footnotesize 16};

\draw[->] (12.5,9.5) -- (13.5,9.5);
\filldraw[black] (13,9.5) circle (0pt) node[anchor=south] {\footnotesize 17};
\filldraw[black] (10.5,9.5) circle (0pt) node[anchor=west] {$T(eR\phi)^{ab}{}_k$};
\filldraw[black] (14,9.5) circle (0pt) node[anchor=west] {$\lambda(\phi)^{ab}{}_{0i}$};

\filldraw[black] (0,8) circle (0pt) node[anchor=west] {$P(B)_{ab}{}^{0i}$};
\draw[->] (2,8) -- (3,8);
\filldraw[black] (2.5,8) circle (0pt) node[anchor=south] {\footnotesize 2};
\filldraw[black] (3.5,8) circle (0pt) node[anchor=west] {$S(R\phi)^{abi}$};
\draw[->] (5.5,8) -- (6.5,8);
\filldraw[black] (6,8) circle (0pt) node[anchor=south] {\footnotesize 13};
\filldraw[black] (7,8) circle (0pt) node[anchor=west] {$\lambda(\phi)_{abij}$};

\filldraw[black] (0,7) circle (0pt) node[anchor=west] {$P(\phi)_{ab}{}^{ij}$};
\draw[->] (2,7) -- (3,7);
\filldraw[black] (2.5,7) circle (0pt) node[anchor=south] {\footnotesize 3};
\filldraw[black] (3.5,7) circle (0pt) node[anchor=west] {$S(Bee)^{abij}$};
\draw[->] (5.5,7) -- (6.5,7);
\filldraw[black] (6,7) circle (0pt) node[anchor=south] {\footnotesize 14};
\filldraw[black] (7,7) circle (0pt) node[anchor=west] {$\lambda(B)_{ab0i}$};

\filldraw[black] (0,6) circle (0pt) node[anchor=west] {$P(B)_{ab}{}^{ij}$};
\draw[->] (2,6) -- (3,6);
\filldraw[black] (2.5,6) circle (0pt) node[anchor=south] {\footnotesize 4};
\filldraw[black] (3.5,6) circle (0pt) node[anchor=west] {$\lambda(\omega)_{abi}$};

\filldraw[black] (0,5) circle (0pt) node[anchor=west] {$P(\beta)_a{}^{ij}$};
\draw[->] (2,5) -- (3,5);
\filldraw[black] (2.5,5) circle (0pt) node[anchor=south] {\footnotesize 5};
\filldraw[black] (3.5,5) circle (0pt) node[anchor=west] {$\lambda(e)_{ai}$};

\filldraw[black] (0,4) circle (0pt) node[anchor=west] {$P(\omega)_{ab}{}^i$};
\draw[->] (2,4) -- (3,4);
\filldraw[black] (2.5,4) circle (0pt) node[anchor=south] {\footnotesize 6};
\filldraw[black] (3.5,4) circle (0pt) node[anchor=west] {$\lambda(B)_{abij}$};

\filldraw[black] (0,3) circle (0pt) node[anchor=west] {$P(\phi)_{ab}{}^{0i}$};
\draw[->] (2,3) -- (3,3);
\filldraw[black] (2.5,3) circle (0pt) node[anchor=south] {\footnotesize 7};
\filldraw[black] (3.5,3) circle (0pt) node[anchor=west] {$S(Bee)^{abi}$};
\draw (5.5,3) -- (6,3);
\draw (6,2) -- (6,3);
\draw[->] (6,2.5) -- (6.8,2.5);
\draw[->] (9,2.5) -- (10,2.5);
\filldraw[black] (6.4,2.5) circle (0pt) node[anchor=south] {\footnotesize 8};
\filldraw[black] (7,2.5) circle (0pt) node[anchor=west] {$T(\beta)^a{}_{\mu\nu}$};
\filldraw[black] (9.5,2.5) circle (0pt) node[anchor=south] {\footnotesize 9};
\filldraw[black] (10.5,2.5) circle (0pt) node[anchor=west] {$\lambda(\beta)^a{}_{\mu\nu}$};
\filldraw[black] (0,2) circle (0pt) node[anchor=west] {$P(\omega)_{ab}{}^0$};
\draw (2,2) -- (6,2);

\filldraw[black] (0,1) circle (0pt) node[anchor=west] {$P(e)_a{}^0$};
\draw[->] (2,1) -- (3,1);
\filldraw[black] (2.5,1) circle (0pt) node[anchor=south] {\footnotesize 10};
\filldraw[black] (3.5,1) circle (0pt) node[anchor=west] {$S(eR)^a$};
\draw[->] (5.5,1) -- (6.5,1);
\filldraw[black] (6,1) circle (0pt) node[anchor=south] {\footnotesize 12};
\filldraw[black] (7,1) circle (0pt) node[anchor=west] {$0$};
\end{tikzpicture}
\end{center}
Here every arrow represents one consistency requirement, and numbers on the arrows denote the order in which we will discuss them. Steps 8 and 16 involve multiple constraints simultaneously, and will require special consideration.
 Primary, secondary and tertiary constraints are denoted as $P$, $S$ and $T$, respectively.

We begin by discussing consistency conditions 1--7,
\begin{equation}
\begin{array}{c}
\dot{P}(\beta)_a{}^{0i} \approx 0\,, \qquad \dot{P}(B)_{ab}{}^{0i} \approx 0\,, \qquad
\dot{P}(\phi)_{ab}{}^{ij} \approx 0\,, \qquad \dot{P}(\phi)_{ab}{}^{0i} \approx 0\,, \\
\dot{P}(B)_{ab}{}^{ij} \approx 0\,, \qquad \dot{P}(\beta)_a{}^{ij} \approx 0\,, \qquad
\dot{P}(\omega)_{ab}{}^i \approx 0\,. \\
\end{array}
\end{equation}
Calculating the corresponding Poisson brackets with the total Hamiltonian, these give rise to the following secondary constraints,
\begin{equation}
\begin{array}{lcl}
S(T)^{ai} & \equiv & \lc^{0ijk} T^a{}_{jk} \approx 0\,, \\
S(R\phi)^{abi} & \equiv & \lc^{0ijk} \left( R^{ab}{}_{jk} - \phi^{ab}{}_{jk} \right) \approx 0\,, \\
S(Bee)^{abij} & \equiv & \lc^{0ijk} \left( B^{ab}{}_{0k} - 2 \lc^{abcd}\, e_{c0} e_{dk} \right) \approx 0\,, \\
S(Bee)^{abi} & \equiv &  \lc^{0ijk} \left( B^{ab}{}_{jk} - 2 \lc^{abcd}\, e_{cj} e_{dk} \right) \approx 0\,, \\
\end{array}
\end{equation}
and determine the following multipliers,
\begin{equation}
\begin{array}{lcl}
\lambda(\omega)^{ab}{}_i & \approx & \ds \nabla_i \omega^{ab}{}_0 + \phi^{ab}{}_{0i} \,, \\
\lambda(e)^a{}_i & \approx & \ds \nabla_i e^a{}_0 -  \omega^a{}_{b0} e^b{}_i \,, \\
\lambda(B)^{ab}{}_{ij} & \approx & \ds 4 \lc^{abcd} \left( \nabla_{[i} e_{c0} - \omega_{cf0} e^f{}_{[i} \right) e_{dj]} + e^{[a}{}_0 \beta^{b]}{}_{ij} - 2 e^{[a}{}_{[i} \beta^{b]}{}_{0j]} \,. \\
\end{array}
\end{equation}
In step 8 we discuss the consistency conditions
\begin{equation}
\dot{S}(Bee)^{abi} \approx 0\,, \qquad \dot{P}(\omega)_{ab}{}^0 \approx 0\,,
\end{equation}
simultaneously. Calculating the time derivatives, we obtain
\begin{equation}
\lc^{0ijk} \left( e^{[a}{}_0 \beta^{b]}{}_{jk} - 2 e^{[a}{}_j \beta^{b]}{}_{0k} \right) \approx 0\,, \qquad \lc^{0ijk} \, e^{[a}{}_i \beta^{b]}{}_{jk} \approx 0\,,
\end{equation}
which can be jointly written as a covariant equation
\begin{equation}
\lc^{\mu\nu\rho\sigma} \, e^{[a}{}_{\nu} \beta^{b]}{}_{\rho\sigma} \approx 0\,.
\end{equation}
With the assumption that $\det (e^a{}_{\mu}) \neq 0$, this can be solved for $\beta^a$, giving a set of very simple tertiary constraints:
\begin{equation} \label{CsTbeta}
T(\beta)^a{}_{\mu\nu} \equiv \beta^a{}_{\mu\nu} \approx 0 \,.
\end{equation}
At this point we can immediately analyze the consistency step 9 as well. Taking the time derivative of (\ref{CsTbeta}), one easily determines the corresponding multipliers,
\begin{equation}
\lambda(\beta)^a{}_{\mu\nu} \approx 0\,.
\end{equation}
Next, in steps 10 and 11, from the consistency conditions for the remaining two primary constraints,
\begin{equation}
\dot{P}(e)_a{}^0 \approx 0\,, \qquad \dot{P}(e)_a{}^i \approx 0\,,
\end{equation}
we obtain two new secondary constraints,
\begin{equation}
\begin{array}{lcl}
S(eR)_a & \equiv & \lc^{0ijk} \lc_{abcd} e^b{}_i R^{cd}{}_{jk} \approx 0\,, \\
S(eR\phi)_a{}^i & \equiv & \ds \lc^{0ijk}\lc_{abcd} \left( e^b{}_0 R^{cd}{}_{jk} - 2 e^b{}_j \phi^{cd}{}_{0k} \right) \approx 0\,. \\
\end{array}
\end{equation}
In step 12 we need to discuss the consistency condition for the constraint $S(eR)_a$. After a straightforward but tedious calculation, one eventually ends up with the following expression:
\begin{equation}
\dot{S}(eR)_a = \nabla_i S(eR\phi)_a{}^i + \omega^b{}_{a0} S(eR)_b + 2\lc_{abcd}\phi^{cd}{}_{0k} S(T)^{bk}\,,
\end{equation}
up to terms proportional to primary constraints. Since the time derivative is already expressed as a linear combination of constraints, the consistency condition is trivially satisfied, which is denoted with a zero in the diagram above.

Moving on to steps 13, 14 and 15, the consistency conditions
\begin{equation}
\dot{S}(R\phi)^{abi} \approx 0\,, \qquad \dot{S}(Bee)^{abij} \approx 0\,, \qquad \dot{S}(T)^{ai} \approx 0\,,
\end{equation}
determine the multipliers
\begin{equation}
\begin{array}{lcl}
\lambda(\phi)^{ab}{}_{jk} & \approx & \ds 2 \omega^{[a}{}_{c0} R^{b]c}{}_{jk} + 2\nabla_{[j} \phi^{ab}{}_{0k]}\,,\\
\lambda(B)_{ab0k} & \approx & \ds 2 \lc_{abcd} \left[ e^d{}_k \lambda(e)^c{}_0 - e^d{}_0 \nabla_k e^c{}_0 + \omega^c{}_{f0} e^d{}_0 e^f{}_k \right] \,, \\
\end{array}
\end{equation}
and another tertiary constraint
\begin{equation}
T(eR\phi)^{ai} \equiv \lc^{0ijk} \left( R^{ab}{}_{jk} e_{b0} + 2\phi^{ab}{}_{0j} e_{bk} \right) \approx 0\,.
\end{equation}
Now we turn to step 16. At this point there are only two constraints, $T(eR\phi)^{ai}$ and $S(eR\phi)^{ai}$, whose consistency conditions have not been discussed yet. To this end, note that these two constraints can be rewritten into a very similar form,
\begin{equation} \label{NezKonstraintiZajedno}
\begin{array}{lcr}
S(eR\phi)_a{}^i & = & \ds  \lc_{abcd} \lc^{0ijk} \left( e^b{}_0 R^{cd}{}_{jk}  - 2 e^b{}_j \phi^{cd}{}_{0k} \right) \,, \\
T(eR\phi)_a{}^i & = & \ds \eta_{ac} \eta_{bd} \lc^{0ijk} \left( e^b{}_0 R^{cd}{}_{jk}  - 2 e^b{}_j \phi^{cd}{}_{0k} \right) \,, \\
\end{array}
\end{equation}
where the identical expression in parentheses is contracted with $\lc_{abcd}$ in the first constraint and with $\eta_{ac} \eta_{bd}$ in the second. This suggests that we should discuss their consistency conditions simultaneously. As suggested in the diagram above, we will first rewrite these $24$ constraints (\ref{NezKonstraintiZajedno}) into a system of $18+6$ constraints (to be denoted $T(eR\phi)_{abk}$ and $T(eR\phi)_{jk}$ respectively) as follows. Given that the tetrad $e^a{}_{\mu}$ is nondegenerate, we can freely multiply the constraints with it and split the index $\mu$ into space and time components. The $\mu=0$ part is
\begin{equation} \label{NezKonstraintiVremenski}
\begin{array}{lcr}
e^a{}_0 S(eR\phi)_a{}^i & = & \ds  - 2 \lc_{abcd} \lc^{0ijk} e^a{}_0 e^b{}_j \phi^{cd}{}_{0k} \,, \\
e^a{}_0 T(eR\phi)_a{}^i & = & \ds  - 2 \eta_{ac} \eta_{bd} \lc^{0ijk} e^a{}_0 e^b{}_j \phi^{cd}{}_{0k} \,, \\
\end{array}
\end{equation}
where the curvature terms have automatically vanished, while the $\mu=m$ part is
\begin{equation} \label{NezKonstraintiProstorni}
\begin{array}{lcr}
e^a{}_m S(eR\phi)_a{}^i & = & \ds e^a{}_m  \lc_{abcd} \lc^{0ijk} \left( e^b{}_0 R^{cd}{}_{jk}  - 2 e^b{}_j \phi^{cd}{}_{0k} \right) \,, \\
e^a{}_m T(eR\phi)_a{}^i & = & \ds e^a{}_m \eta_{ac} \eta_{bd} \lc^{0ijk} \left( e^b{}_0 R^{cd}{}_{jk}  - 2 e^b{}_j \phi^{cd}{}_{0k} \right) \,. \\
\end{array}
\end{equation}
The system of $18$ constraints (\ref{NezKonstraintiProstorni}) can be shown to be equivalent to the following constraint:
\begin{equation} \label{ResenaVezaZaPhi}
T(eR\phi)^{ab}{}_k \equiv \phi^{ab}{}_{0k} - e^f{}_0 R^{cd}{}_{ij} F^{abij}{}_{fcdk}\,,
\end{equation}
where $F^{abij}{}_{fcdk}$ is a complicated function of $e^a{}_i$ only. The proof that the system (\ref{NezKonstraintiProstorni}) is equivalent to (\ref{ResenaVezaZaPhi}) is given in Appendix \ref{AppSistemJna}, and the explicit expression for $F^{abij}{}_{fcdk}$ is given in equation (\ref{KonacnaJnaZaVelikoF}).  Second, introducing the shorthand notation $K_{abcd} \in \{ \lc_{abcd}, \eta_{ac} \eta_{bd} \}$ and using (\ref{ResenaVezaZaPhi}), we define
\begin{equation} \label{KernelJne}
T(eR\phi)^i \equiv -2 K_{abcd} \lc^{0ijk} e^a{}_0 e^b{}_j e^f{}_0 R^{gh}{}_{mn} F^{cdmn}{}_{fghk} \,,
\end{equation}
which represents a set of $3+3=6$ constraints equivalent to (\ref{NezKonstraintiVremenski}). However, a straightforward and meticulous (albeit very long) calculation shows that the expression (\ref{KernelJne}) is already a linear combination of known constraints and Bianchi identities, and is thus already weakly equal to zero. Therefore, $T(eR\phi)^i $ is not a new independent constraint, and its consistency condition is automatically satisfied.

Summing up the step $16$, we have replaced the set of constraints (\ref{NezKonstraintiZajedno}) by an equivalent set (\ref{ResenaVezaZaPhi}). It thus follows that the consistency conditions for $S(eR\phi)_a{}^i$ and $T(eR\phi)_a{}^i$ are equivalent to the consistency condition for $T(eR\phi)^{ab}{}_k$. Consequently, in step $17$, we find that the consistency condition
\begin{equation}
\dot{T}(eR\phi)^{ab}{}_k \approx 0
\end{equation} 
determines the multiplier $\lambda(\phi)^{ab}{}_{0k}$ as
\begin{equation}
\lambda(\phi)^{ab}{}_{0k} \approx \lambda(e)^f{}_0 R^{cd}{}_{ij} F^{abij}{}_{fcdk} +
 2 e^f{}_0 \left[ R^c{}_{hij} \omega^{hd}{}_0 + \nabla_i \phi^{cd}{}_{0j} \right] F^{abij}{}_{fcdk} +
 e^f{}_0 R^{cd}{}_{ij} \frac{\del F^{abij}{}_{fcdk}}{\del e^h{}_m} \left( \nabla_m e^h{}_0 - \omega^h{}_{g0} e^g{}_m  \right) \,.
\end{equation}
This concludes the consistency procedure for all constraints.

\subsection{Results}

Let us sum up the results of the consistency procedure. We have determined the full set of constraints and multipliers as follows: the primary constraints are
\begin{equation}
P(B)_{ab}{}^{\mu\nu}\,, \qquad
P(\phi)_{ab}{}^{\mu\nu}\,, \qquad
P(\beta)_a{}^{\mu\nu}\,, \qquad
P(\omega)_{ab}{}^{\mu}\,, \qquad
P(e)_a{}^{\mu}\,,
\end{equation}
and they have $36$, $36$, $24$, $24$ and $16$ components, respectively, or $136$ in total. The secondary constraints are
\begin{equation}
S(T)^{ai}\,, \qquad
S(R\phi)^{abi}\,, \qquad
S(Bee)^{abij}\,, \qquad
S(Bee)^{abi}\,, \qquad
S(eR)^{a}\,,
\end{equation}
and they have $12+18+18+18+4=70 $ components in total. The tertiary constraints are
\begin{equation}
T(\beta)^a{}_{\mu\nu}\,, \qquad
T(eR\phi)^{ab}{}_i\,.
\end{equation}
and they have $24+18=42$ components. In addition, the determined multipliers are
\begin{equation}
\lambda(B)^{ab}{}_{\mu\nu}\,, \qquad
\lambda(\phi)^{ab}{}_{\mu\nu}\,, \qquad
\lambda(\beta)^a{}_{\mu\nu}\,, \qquad
\lambda(\omega)^{ab}{}_i\,, \qquad
\lambda(e)^a{}_i\,,
\end{equation}
and they have $36+36+24+18+12=126 $ components. Finally, there are $10$ remaining undetermined multipliers,
\begin{equation}
\lambda(\omega)^{ab}{}_0\,, \qquad
\lambda(e)^a{}_0\,.
\end{equation}
In total, there are $C= 136+70+42 = 248$ constraints, $126$ determined and $10$ undetermined multipliers, the latter corresponding to the $10$ parameters of the local Poincar\'e symmetry of the action.

\bigskip

\section{\label{SecIV}The physical degrees of freedom}

Once we have found all the constraints in the theory, we need to classify each constraint as a first-class or a second-class constraint. While some of the second class constraints can be identified from (\ref{AlgebraPrimarnihVeza}), the classification is not easy since constraints are unique only up to linear combinations. The most efficient way to tabulate all first class constraints is to substitute all determined multipliers into the total Hamiltonian (\ref{TotalniHamiltonijan}) and rewrite it in the form
\begin{equation} \label{TotHamiltonijanKombVezaPrveKlase}
H_T = \int \rmd^3\vec{x} \left[ \frac{1}{2} \lambda(\omega)^{ab}{}_0 \,\Phi(\omega)_{ab}  + \lambda(e)^a{}_0 \,\Phi(e)_a + \frac{1}{2}\omega^{ab}{}_0 \,\Phi(T)_{ab} + e^a{}_0 \,\Phi(R)_a \right]\,.
\end{equation}
The quantities $\Phi$ are linear combinations of the constraints, and they must all be of the first class, since the total Hamiltonian weakly commutes with all constraints. Written in terms of the primary and the secondary constraints, the first-class constraints are given by
\begin{equation} \label{VezePrveKlaseEksplicitno}
\begin{array}{lcl}
\Phi(\omega)^{ab} & = & \ds P(\omega)^{ab0}\,, \vphantom{\ds\int} \\
\Phi(e)_a & = & \ds P(e)_a{}^0 + \frac{1}{2} R^{cd}{}_{ij} F^{fbij}{}_{acdk} P(\phi)_{fb}{}^{0k} + \lc_{abcd} e^b{}_k P(B)^{cd0k} \,, \vphantom{\ds\int} \\
\Phi(T)^{ab}\!\! & = & \ds 4 \lc^{abcd} e_{ci} S(T)_d{}^i -\nabla_i S(Bee)^{abi} + \lc^{0ijk} e^{[a}{}_i T(\beta)^{b]}{}_{jk} + 2\lc^{abcd} e^f{}_i e_{cj} P(B)_{fd}{}^{ij} \vphantom{\ds\int} \\
 & & \ds  - \nabla_i P(\omega)^{abi} + 2 e^{[a}{}_i P(e)^{b]i} - R^{[ac}{}_{ij} P(\phi)_c{}^{b]ij} \,,  \vphantom{\ds\int}  \\
\Phi(R)_a & = & \ds -S(eR)_a + R^c{}_{hij} \omega^{hd}{}_0 F^{fbij}{}_{acdk} P(\phi)_{fb}{}^{0k} + R^{cd}{}_{ij} \frac{\del F^{fbij}{}_{acdk}}{\del e^h{}_m} \left( \nabla_m e^h{}_0 - \omega^h{}_{g0} e^g{}_m \right) P(\phi)_{fb}{}^{0k} \vphantom{\ds\int} \\
 & & \ds + \frac{1}{2} R^{cd}{}_{ij} F^{fbij}{}_{acdk} \left[ S(Bee)_{fb}{}^k + P(\omega)_{fb}{}^k + \nabla_m P(\phi)_{fb}{}^{km} -2\nabla_m \left( e^e{}_0 F^{ghmk}{}_{efbn} P(\phi)_{gh}{}^{0n} \right) \right] \vphantom{\ds\int} \\
 & & \ds - \lc^{0ijk} \nabla_i T(\beta)_{ajk} + \lc_{abcd} e^b{}_i \nabla_j P(B)^{cdij} -\nabla_i P(e)_a{}^i +\lc_{abcd} \left( \nabla_k e^b{}_0 - \omega^b{}_{f0} e^f{}_k \right) P(B)^{cd0k} \,. \vphantom{\ds\int} \\
\end{array}
\end{equation}

The constraints (\ref{VezePrveKlaseEksplicitno}) are the first-class constraints in the theory. The remaining constraints are of the second class
\begin{equation} \label{VezeDrugeKlaseEksplicitno}
\begin{array}{lcl}
\chi(B)_{ab}{}^{\mu\nu} & = & P(B)_{ab}{}^{\mu\nu}\,, \\
\chi(\phi)_{ab}{}^{\mu\nu} & = & P(\phi)_{ab}{}^{\mu\nu}\,, \\
\chi(\beta)_a{}^{\mu\nu} & = & P(\beta)_a{}^{\mu\nu}\,, \\
\chi(\omega)_{ab}{}^i & = & P(\omega)_{ab}{}^i\,, \\
\chi(e)_a{}^i & = & P(e)_a{}^i\,, \\
\end{array}
\qquad
\begin{array}{lcl}
\chi(T)^{ai} & = & S(T)^{ai}\,, \\
\chi(R\phi)^{abi} & = & S(R\phi)^{abi}\,, \\
\chi(Bee)^{abij} & = & S(Bee)^{abij}\,, \\
\chi(Bee)^{abi} & = & S(Bee)^{abi}\,, \\
\end{array}
\qquad
\begin{array}{lcl}
\chi(\beta)^a{}_{\mu\nu} & = & T(\beta)^a{}_{\mu\nu}\,, \\
\chi(eR\phi)^{ab}{}_i & = & T(eR\phi)^{ab}{}_i\,. \\
\end{array}
\end{equation}

Note that $\chi(\beta)_a{}^{\mu\nu}$ and $\chi(\beta)^a{}_{\mu\nu}$ are different constraints, despite similar notation. Of course, there is no possibility of confusion since we will never raise or lower spacetime indices of these constraints in the rest of this paper. Also, note that despite the fact that there are $12$ components of $\chi(T)^{ai}$, only $6$ of them can be considered second class, since the other $6$ are part of the first class constraint $\Phi(T)^{ab}$.

At this point we can count the physical degrees of freedom. Given a field theory with $N$ fields whose canonical formulation possesses $F$ first-class constraints, one can gauge fix $F$ fields. The second-class constraints do not generate any gauge symmetries and $S$ second-class constraints are equivalent to vanishing of $S/2$ fields and $S/2$ canonically conjugate momenta. Hence the number of independent (physical) fields is given by
\begin{equation} \label{JnaZaBrojStepeniSlobode}
n = N - F - \frac{S}{2}\, .
\end{equation}

The number of field components for each of the fundamental fields is
$$
\begin{array}{|c|c|c|c|c|} \hline
\omega^{ab}{}_{\mu} & \beta^a{}_{\mu\nu} & e^a{}_{\mu} & B^{ab}{}_{\mu\nu} & \phi^{ab}{}_{\mu\nu} \\ \hline
24 & 24 & 16 & 36 & 36 \\ \hline
\end{array}
$$
which gives the total $N=136$. The number of components of the first class constraints is
$$
\begin{array}{|c|c|c|c|c|c|c|c|} \hline
 \Phi(e)_a & \Phi(\omega)_{ab} & \Phi(R)^a & \Phi(T)^{ab} \\ \hline
 4 & 6 & 4 & 6 \\ \hline
\end{array}
$$
which gives the total of $F=20$. Similarly, the number of components for the second class constraints is
$$
\begin{array}{|c|c|c|c|c|c|c|c|c|c|c|} \hline
\chi(B)_{ab}{}^{\mu\nu} &
\chi(\phi)_{ab}{}^{\mu\nu} &
\chi(\beta)_a{}^{\mu\nu} &
\chi(\omega)_{ab}{}^i &
\chi(e)_a{}^i &
\chi(T)^{ai} &
\chi(R\phi)^{abi} &
\chi(Bee)^{abij} &
\chi(Bee)^{abi} &
\chi(\beta)^a{}_{\mu\nu} &
\chi(eR\phi)^{ab}{}_i \\ \hline
 36 & 36 & 24 & 18 & 12 & 12-6 & 18 & 18 & 18 & 24 & 18 \\ \hline
\end{array}
$$
where we have denoted that only $6$ of the total $12$ components of $\chi(T)^{ai}$ are independent. Thus the total number of independent second class constraints is $S=228$. This number can also be deduced as the difference between the previously counted total number of constraints $C=248$ and the number of first class constraints $F=20$.

Finally, substituting $N$, $F$ and $S$ into (\ref{JnaZaBrojStepeniSlobode}), we obtain:
\begin{equation} \label{BrojFizickihStepeniSlobode}
n = 136 - 20 - \frac{228}{2} = 2\,.
\end{equation}
We conclude that the theory has two physical degrees of freedom, as expected for general relativity.

At this point it is convenient to rewrite the last term in (\ref{TotHamiltonijanKombVezaPrveKlase}) in the traditional ADM form. This is done by projecting the constraint $\Phi(R)_a$ onto the hypersurface $\Sigma$ and its orthogonal direction. Using the inverse tetrad $e^{\mu}{}_a$, define the unit vector $n_a$ orthogonal to $\Sigma$ as
\begin{equation}
n_a \equiv \frac{e^0{}_a}{\sqrt{-g^{00}}}
\end{equation}
where $g^{00} \equiv \eta^{ab} e^0{}_a e^0{}_b$ is the time-time component of the inverse metric $g^{\mu\nu}$. The vector $n_a$ is thus normalized, $n_a n^a = -1$, and we can define the orthogonal and parallel projectors with respect to $\Sigma$ as
\begin{equation}
P_{\orto}^a{}_b \equiv - n^a n_b\,, \qquad P_{\para}^a{}_b \equiv \delta^a_b + n^a n_b\,.
\end{equation}
One can then employ these projectors to rewrite the final term in (\ref{TotHamiltonijanKombVezaPrveKlase}) as
\begin{equation}
\begin{array}{lcl}
e^a{}_0 \Phi(R)_a & = & e^a{}_0 \left( P_{\orto}^b{}_a + P_{\para}^b{}_a \right) \Phi(R)_b \\
 & = & -e^a{}_0 n_a n^b \Phi(R)_b + e^a{}_0 P_{\para}^b{}_a \left( e^{\mu}{}_b e^c{}_{\mu} \right) \Phi(R)_c \\
 & = & \left[ e^a{}_0 n_a \vphantom{P_{\para}^b} \right] \left[ - n^b \Phi(R)_b \vphantom{P_{\para}^b} \right] + \left[ e^a{}_0 P_{\para}^b{}_a e^i{}_b \vphantom{P_{\para}^b} \right] \left[ e^c{}_i \Phi(R)_c \vphantom{P_{\para}^b} \right] + \left[ e^a{}_0 P_{\para}^b{}_a e^0{}_b \vphantom{P_{\para}^b} \right] \left[ e^c{}_0 \Phi(R)_c \vphantom{P_{\para}^b} \right] \\
 & = & N \cH_{\orto} + N^i \cD_i\,. \\
\end{array}
\end{equation}
Note that the final term in the second-to-last equality drops out because $P_{\para}^b{}_a e^0{}_b = \sqrt{-g^{00}} P_{\para}^b{}_a n_b \equiv 0$. In the last equality we have introduced the well known ADM lapse and shift functions,
\begin{equation}
N \equiv e^a{}_0 n_a = \frac{1}{\sqrt{-g^{00}}}\,, \qquad N^i \equiv e^a{}_0 P_{\para}^b{}_a e^i{}_b = - \frac{g^{0i}}{g^{00}}\,,
\end{equation}
and we have split the constraint $\Phi(R)_a$ into the scalar constraint and $3$-diffeomorphism constraint,
\begin{equation}
\cH_{\orto} \equiv - n^b \Phi(R)_b \,, \qquad \cD_i \equiv e^c{}_i \Phi(R)_c \,.
\end{equation}

The constraints $\Phi(T)^{ab}$ are equivalent to the local Lorentz constraints $\cJ^{ab}$, which generate the local Lorentz transformations, and together with the $10$ momentum constraints $\Phi(\omega)^{ab}$ and $\Phi(e)_a$, one can use the scalar constraint $\cH_{\orto}$ and the $3$-diffeomorphism constraint $\cD_i$ to find the Poisson bracket algebra of the first-class constraints. This algebra takes the form
\begin{equation} \label{AlgebraVezaPrveKlase}
\begin{array}{lcl}
\pb{\cJ^{ab}(x)}{\cJ^{cd}(y)} & = & \ds \frac{1}{2} \Big[ \eta^{a[c} \cJ^{d]b}(x) - \eta^{b[c} \cJ^{d]a}(x) \Big] \,\delta^{(3)} \,, \vphantom{\ds\frac{1}{2}} \\
\pb{\cD_{i}(x)}{\cD_{j}(y)} & = & \ds \Big[ \cD_i(x) + \cD_i(y)\Big]\, \partial_j \delta^{(3)} + R^{ab}{}_{ij}(x) \cJ_{ab}(x) \,\delta^{(3)} \,, \vphantom{\ds\frac{1}{2}} \\
\pb{\cD_{i}(x)}{\cH_{\orto}(y)} & = & \ds  \Big[ \cH_{\orto}(x)+ \cH_{\orto}(y)\Big]\,\partial_i \delta^{(3)} + R^{ab}{}_{i0}(x) \cJ_{ab}(x) \, \delta^{(3)} \,, \vphantom{\ds\frac{1}{2}} \\
\pb{\cH_{\orto}(x)}{\cH_{\orto}(y)} & = & \ds \Big[ \tg^{ij}(x) D_j(x)+ \tg^{ij}(y) D_j(y) \Big]\,\partial_i \delta^{(3)} \,, \vphantom{\ds\frac{1}{2}} \\
\end{array}
\end{equation}
while all other first-class Poisson brackets are zero, see \cite{B2}. Here it is assumed that $x\equiv (t,\vec{x})$, $y\equiv (t,\vec{y})$ and $\delta^{(3)} \equiv \delta^{(3)}(\vec{x}-\vec{y})$, while $\tg^{ij}$ is the $3D$ inverse metric, defined in Appendix \ref{AppC}.

\bigskip
The Poisson brackets between the second class constraints and the Poisson brackets between the first and the second class constraints can be calculated, but we do not give their explicit form because we do not need these Poisson brackets for the purposes of this paper. Their generic structure is given by
\begin{equation} \label{AlgebraVezaDrugeKlase}
\pb{\chi_I (x)}{\chi_J (y)} = \Delta_{IJ}(x,y) + \tilde\Delta_{IJ}(x,y) \,,
\end{equation}
and
\begin{equation} \label{AlgebraMesovitihVeza}
\pb{\Phi_A (x)}{\chi_I (y)} = f_{AI}{}^B(x,y)\,\Phi_B (x) +\tilde f_{AI}{}^B(x,y)\,\Phi_B (y) + f_{AI}{}^J(x,y) \,\chi_J (x) +\tilde f_{AI}{}^J(x,y) \,\chi_J (y)\,.
\end{equation}
If we denote all the fields collectively as $\theta^N = ( e^a{}{}_{\mu}, \omega^{ab}{}_{\mu}, \beta^a{}_{\mu\nu}, B^{ab}{}_{\mu\nu}, \phi^{ab}{}_{\mu\nu} )$ and their corresponding momenta as $\pi_N = ( \pi(e)_a{}{}^{\mu}, \pi(\omega)_{ab}{}^{\mu}, \pi(\beta)_a{}^{\mu\nu}, \pi(B)_{ab}{}^{\mu\nu}, \pi(\phi)_{ab}{}^{\mu\nu} ) $, we can denote $\Delta$ and $f$ as generalized functions of the type
$$
F(\theta(x),\pi(x))\delta^{(3)} + F^i(\theta(x),\pi(x))\,\partial_i\delta^{(3)} + \cdots
$$
so that all the coefficients are evaluated at the point $x$, while $\tilde\Delta$ and $\tilde f$ as
$$
F(\theta(y),\pi(y))\delta^{(3)} + F^i(\theta(y),\pi(y))\,\partial_i\delta^{(3)} + \cdots
$$
so that all the coefficients are evaluated at the point $y$.

\section{\label{SecV}The phase space reductions}

The results of the Hamiltonian analysis imply that the $BFCG$ GR action (\ref{BFCGdejstvo}) can be written as
\begin{equation}
S_0 = \int_{t_1}^{t_2} dt \int_\Sigma d^3 x \left[ \pi_N \dot{\theta}^N - \lambda(e)^a{}_0 \Phi(e)_a - \frac{1}{2} \lambda(\omega)^{ab}{}_0 \Phi(\omega)_{ab} - e^a{}_0 \Phi(R)_a - \frac{1}{2} \omega^{ab}{}_0 \Phi(T)_{ab} - \mu^L \chi_L \right]\,,
\end{equation}
where $\chi_L$ counts over the set of all second-class constraints (\ref{VezeDrugeKlaseEksplicitno}), while $\mu^L$ are Lagrange multipliers for the second-class constraints.

This action can be reduced to an action for a smaller number of canonical variables by partially solving some of the constraints. Solving $M$ first-class constraints $\phi_m =0$ requires that we make $M$ gauge-fixing conditions $G_m =0$, such that $\{G_m,G_{m'}\} =0$ and $\det\{G_m,\phi_{m'}\} \ne 0$. We can then solve the equations $\phi_m = 0$ for the momenta $\pi(G_m)$. The simplest way to do this is to chose $G_m$ to be a set of $M$ coordinates $\theta_m$, and then to solve the corresponding $M$ first-class constraints $\phi_m =0$ for the momenta $\pi_m$. As far as the second-class constraints are concerned, we can solve $2K$ of them for $K$ coordinates  and their $K$ momenta.

It is not difficult to see that one can solve the following $192$ second-class constraints 
\begin{equation}
\begin{array}{lclcl}
\chi(B)_{ab}{}^{\mu\nu}  & \equiv & \ds \pi(B)_{ab}{}^{\mu\nu} & \approx \vphantom{\ds\int} & 0 \,, \\
\chi(\phi)_{ab}{}^{\mu\nu} & \equiv & \ds \pi(\phi)_{ab}{}^{\mu\nu} & \approx \vphantom{\ds\int} & 0 \,, \\
\chi(\beta)^a{}_{\mu\nu} & \equiv & \ds \beta^a{}_{\mu\nu} & \approx \vphantom{\ds\int} & 0 \,, \\
\chi(\beta)_a{}^{\mu\nu} & \equiv & \ds \pi(\beta)_a{}^{\mu\nu} + \lc^{0\mu\nu\rho} e_{a\rho} & \approx \vphantom{\ds\int} & 0 \,, \\
\end{array}
\qquad\qquad
\begin{array}{lclcl}
\chi(Bee)^{abij} & \equiv & \ds \lc^{0ijk} \left( B^{ab}{}_{0k} - 2 \lc^{abcd} e_{c0} e_{dk} \right) & \approx \vphantom{\ds\int} & 0 \,, \\
\chi(Bee)^{abi} & \equiv & \ds \lc^{0ijk}  \left( B^{ab}{}_{jk} - 2 \lc^{abcd} e_{cj} e_{dk} \right) & \approx \vphantom{\ds\int} & 0 \,, \\
\chi(R\phi)^{abi} & \equiv & \ds \lc^{0ijk} \left( R^{ab}{}_{jk} - \phi^{ab}{}_{jk} \right) & \approx \vphantom{\ds\int} & 0 \,, \\
\chi(eR\phi)^{ab}{}_i & \equiv & \ds \phi^{ab}{}_{0i} - e^f{}_0 R^{cd}{}_{jk} F^{abjk}{}_{fcdi} & \approx \vphantom{\ds\int} & 0 \,, \\
\end{array}
\end{equation}
for $(B,\beta,\phi)$ and their momenta. This will give $(B,\beta,\phi)$ and their momenta as functions of the canonical coordinates $(e,\omega, \pi(e),\pi(\omega))$ so that one obtains a reduced phase-space (RPS) theory described by the action
\begin{equation}
S_1 = \int d^4 x \left[ \pi(e)_a{}^\mu\,\dot{e}^a{}_\mu + \frac{1}{2} \pi(\omega)_{ab}{}^{\mu} \,\dot{\omega}^{ab}{}_{\mu} - \lambda(e)^a{}_0 \tilde{\Phi}(e)_a - \frac{1}{2} \lambda(\omega)^{ab}{}_0 \tilde{\Phi}(\omega)_{ab} - e^a{}_0 \tilde{\Phi}(R)_a - \frac{1}{2} \omega^{ab}{}_0 \tilde{\Phi}(T)_{ab} - \mu^L \tilde\chi_L \right]\,, 
\end{equation}
where $\tilde C$ denotes a constraint $C$ on the RPS $(e,\omega, \pi(e),\pi(\omega))$. There are still $20$ first-class constraints, namely $\tilde{\Phi}(\omega)^{ab}$, $\tilde{\Phi}(e)_a$, $\tilde{\Phi}(T)^{ab}$, $\tilde{\Phi}(R)_a$, and $36$ second-class constraints $\tilde{\chi}_L = (\tilde{\chi}(e)_a{}^i, \tilde{\chi}(\omega)_{ab}{}^i, \tilde{\chi}(T)^{ai})$ on the RPS, so that $S_1$ is equivalent to the Hamiltonian form of the Einstein-Cartan action \cite{B}.

One would like to understand a reduction of $S_1$ to an action for the triads and spatial spin connections $( e^\alpha{}_i , \omega^{\alpha\beta}{}_i )$. This can be done by gauge fixing $e^a{}_0 = 0$ and solving the corresponding momenta from $\tilde{\Phi}(e)_a = 0$. One can also gauge fix $\omega^{ab}{}_0 =0$ and eliminate the corresponding momenta from $\tilde{\Phi}(\omega)^{ab} = 0$, as well as  to set $e^{\ubar 0}{}_i = 0$ and eliminate the corresponding momenta from $\tilde{\Phi}(T)^{\ubar{0}\alpha}=0$. Note that here we have split the group indices into space and time components, $a = (\ubar{0},\alpha)$ where $\alpha=1,2,3$, see Appendix \ref{AppC} for details and the notation.

As far as the second-class constraints $\tilde\chi_L$ are concerned, one can eliminate $\omega^{\ubar 0\,\alpha}{}_i$ and the corresponding momenta from 
\begin{equation}
\tilde\chi(\omega)_{\ubar 0\alpha}{}^i = 0\,,\qquad \tilde{\chi}(e)_a{}^i = 0\,,\qquad \tilde{\chi}(T)^{\ubar{0}i}= 0\,.\label{2cEC}
\end{equation} 
Note that there are 24 constraints in (\ref{2cEC}), but there are 6 relations among them, so that we have only 18 independent constraints.

Solving the constraints (\ref{2cEC}) leads to a RPS based on $(e^{\alpha}{}_i , \omega^{\alpha\beta}{}_i) \cong (e^{\alpha}{}_i , \omega^{\alpha}{}_i)$ and their momenta. However, there are still $7$ first-class constraints
\begin{equation}
\tilde{\Phi}(R)^a = 0\,,\qquad \tilde{\Phi}(T)^{\alpha\beta}=0\,,
\end{equation}
and $18$ second-class constraints 
\begin{equation}
\tilde{\chi}(T)^{\alpha i}=0\,, \qquad \tilde{\chi}(\omega)_{\alpha\beta}{}^i =0 \,.\label{eo2c}
\end{equation} 
The corresponding action is given by 
\begin{equation}
S_2 = \int d^4 x \left[ \pi(e)_{\alpha}{}^i \dot{e}^{\alpha}{}_i + \pi(\omega)_{\alpha}{}^i \dot{\omega}^{\alpha}{}_i - N \tilde{\cH}_{\orto} - N^i \tilde{\cD}_i - \frac{1}{2} \omega^{\alpha\beta}{}_0 \tilde{\cJ}_{\alpha\beta} - \mu^L \tilde\chi_L \right]\,,
\end{equation}
where $\tilde{\chi}_L = (\tilde{\chi}(T)^{\alpha i}, \tilde{\chi}(\omega)_{\alpha\beta}{}^i)$ and $\omega^{\alpha}{}_i \equiv \frac{1}{2} \lc^{\alpha\beta\gamma} \omega_{\beta\gamma i}$.

We can further eliminate $\omega^{\alpha}{}_i$ and their momenta from the $18$ second-class constraints (\ref{eo2c}) so that one obtains a RPS based on $(e,\pi(e))$ variables and the action
\begin{equation}
S_3 = \int d^4 x \left[ \pi(e)_{\alpha}{}^i \dot{e}^{\alpha}{}_i - N \tilde{\cH}_{\orto} - N^i \tilde{\cD}_i - \frac{1}{2} \omega^{\alpha\beta}{}_0 \tilde{\cJ}_{\alpha\beta} \right]\,.
\end{equation}
This action corresponds to the triad Hamiltonian formulation of general relativity. The ADM formulation is obtained by using the $3D$ metric $g_{ij} \equiv e^a{}_i e_{aj} =  e^{\alpha}{}_i e_{\alpha j}$ and the corresponding momenta. The ADM variables are invariant under the local rotations generated by $\tilde{\cJ}^{\alpha\beta}$, so that the corresponding action is given by
\begin{equation}
S_4 = \int d^4 x \left[ \pi(g)^{ij}\,\dot g_{ij} - N \tilde{\cH}_{\orto} - N^i \tilde{\cD}_i \right]\,,
\end{equation}
where $\cH_{\orto}$ and $\cD_i$ are the ADM constraints.

\section{\label{SecVI}Conclusions}

We found all the constraints and determined the Lagrange multipliers for the $BFCG$ GR action (\ref{BFCGdejstvo}). We also determined the total Hamiltonian (\ref{TotHamiltonijanKombVezaPrveKlase}), the first-class constraints (\ref{VezePrveKlaseEksplicitno}), the second-class constraints (\ref{VezeDrugeKlaseEksplicitno}) and the algebra of the constraints (\ref{AlgebraVezaPrveKlase}), (\ref{AlgebraVezaDrugeKlase}) and (\ref{AlgebraMesovitihVeza}). The obtained constraints also give the correct number of the physical DOF, see (\ref{BrojFizickihStepeniSlobode}). We also showed how the other known canonical formulations of GR, namely Einstein-Cartan, triad and ADM, arise from the canonical formulation of $BFCG$ GR by performing the RPS analysis. This analysis also gave a new canonical formulation for GR, namely the action $S_2$, which is based on the reduced phase space of triads and $SO(3)$ connections and their canonically conjugate momenta.

%Of course, all obtained results agree with the Hamiltonian structure expected for general relativity. In that sense, this work does not contain any surprising new results. Nevertheless, given that the $BFCG$ GR action features many auxiliary variables and features the basic structure of the topological $BFCG$ action, it is actually quite important to have explicit expressions for all constraints in the theory, especially if one is interested in the canonical quantization of the model. In this sense, this work represents the culmination of the previous studies of the $BFCG$ theory in terms of its canonical structure \cite{MO,MOV,MOVdrugi}.

%Regarding the future lines of research, by far the most important topic is tha canonical quantization programme of the $BFCG$ GR action. Namely, the covariant quantization of the theory has been studied \cite{MV,M,Vojinovic} (see also \cite{BaratinFreidel2015}), revealing the structure of the Regge quantum gravity (RQG) model. The corresponding canonical quantization can resolve the long-standing open problem of the canonical structure for RQG. The aim is to find a categorical generalization of the spin-network basis, that would be compatible with RQG.

Since the main motivation for finding a canonical formulation of the $BFCG$ GR theory is the construction of a spin-foam basis which will be a categorical generalization of the spin-network basis from LQG, then the results of the RPS analysis in section 5 are of great importance for this goal. Namely, in order to construct such a spin-foam basis one needs a 2-connection $(A,\beta)$  for the Euclidean 2-group $(SO(3),\realni^3 )$ on the spatial manifold $\Sigma$, see \cite{MO}. This makes the RPS space $(e^\alpha{}_i,\omega^{\alpha\beta}{}_i,\pi(e)_\alpha{}^i,\pi(\omega)_{\alpha\beta}{}^i)$ and the corresponding action $S_2$ a natural starting point for the canonical quantization. Furthermore, this RPS provides a natural 2-connection on $\Sigma$
\begin{equation}
(A^{\alpha\beta}{}_i\, , \beta^\alpha{}_{ij} ) = ( \omega^{\alpha\beta}{}_i \,, \epsilon_{ijk}\teu{k}{\alpha} )\,,\label{2conn}
\end{equation}
where $\te{k}{\alpha}$ are the inverse triads.

Hence one can use the 2-holonomy invariants for the 2-connection (\ref{2conn}) associated to embedded 2-graphs in $\Sigma$, see \cite{fmr}, in order to construct the wavefunctions corresponding to the spin-foam basis. However, the existence of the second-class constraints $\chi_m$ will complicate the task of obtaining the physical Hilbert space. One can avoid the second-class constraints by using the Dirac brackets, but this may produce non-canonical commutators among the fields and their canonical momenta. If one wants to preserve the Heisenberg algebra of the canonical variables, then one can use the Gupta-Bleuler quantization approach, where the second-class constraints would be imposed weakly, as $\langle \Psi|\hat\chi_m |\Psi \rangle =0$.

A simpler approach to the problem of second-class constraints in quantum theory is to solve classically the second-class constraints $\chi_m$, which is equivalent to using the $(e^\alpha{}_i ,\pi(e)_\alpha{}^i)$ RPS and the action $S_3$. Then the spin connection $\omega^{\alpha\beta}{}_i$ becomes a function of the triads and the components of the 2-connection (\ref{2conn}) will still commute as operators, so that a spin-foam basis can be constructed, and the $e$-representation will be the most convenient for this. 

Note that in the triad formulation of GR the Ashtekar variables can be defined via a series of canonical transformations, 
\begin{equation}
(e^{\alpha}{}_i , \pi(e)_{\alpha}{}^i ) \to ( \te{i}{\alpha} , \pi(\tilde e)_i{}^{\alpha} ) \to \Big(f(\zeta) E^i{}_{\alpha} = |e|_3\,\te{i}{\alpha} \,,\, A^{\alpha}{}_i = \omega(e)^\alpha{}_i + \frac{\zeta}{|e|_3} \pi(\tilde e)_i{}^\alpha \Big) \,,
\end{equation}
where $|e|_3 = \det(e^\alpha{}{}_i)$ and for $\zeta =\sqrt{-1}$, $f(\zeta)=1$ \cite{A} while for $\zeta\in\bf \realni$, $f(\zeta)=\zeta$ \cite{Ba}. Then one can define the spin-network basis by using spin-network graphs and the associated holonomies for the connection $A$, see \cite{lqg}. This suggests that the Ashtekar variables could be also a natural starting point for the construction of a spin-foam basis. However, the corresponding 2-connection components
\begin{equation}
A^{\alpha\beta}{}_i =\epsilon^{\alpha\beta\gamma} A_{\gamma i} \,, \qquad\beta^{\alpha}{}_{ij} = \epsilon_{ijk} E^{k\alpha} \,,
\end{equation}
will not commute as operators and one has to use again the 2-connection (\ref{2conn}).  

Let us also note that the results obtained about the Hamiltonian structure of the theory can be important if one considers minisuperspace or midisuperspace models of quantum gravity, as is commonly done in the context of cosmology. For example, in Loop Quantum Cosmology (for a review, see \cite{AshtekarSingh,Singh,MartinBenito,WilsonEwing} and references therein), one typically performs some type of symmetry reduction or gauge fixing prior to quantization, and then considers a resulting quantum-mechanical model of the Universe. However, in this work we have discussed only pure gravity, without matter fields. For this reason, our results are not directly applicable in the context of cosmology, since cosmological models without matter fields are not realistic. Repeating our analysis with included matter fields therefore represents an interesting avenue for further research.

\begin{acknowledgments}
MV would like to thank Prof. Milovan Vasili\'c for discussion and helpful suggestions. AM was partially supported by the FCT projects PEst-OE/MAT/UI0208/2013 and EXCL/MAT-GEO/0222/2012. MAO was supported by the FCT grant SFRH/BD/79285/2011. MV was supported by the FCT project PEst-OE/MAT/UI0208/2013, the FCT grant SFRH/BPD/46376/2008, the bilateral project ``Quantum Gravity and Quantum Integrable Models - 2015-2016'' number 451-03-01765/2014-09/24 between Portugal and Serbia, and partially by the project ON171031 of the Ministry of Education, Science and Technological Development, Serbia.
\end{acknowledgments}

\appendix

\section{Bianchi identities}
\setcounter{section}{1}

Recalling the definitions of the torsion and curvature $2$-forms,
\begin{equation}
T^a = de^a + \omega^a{}_b \wedge e^b\,, \qquad R^{ab} = d\omega^{ab} + \omega^a{}_c \wedge \omega^{cb}\,,
\end{equation}
one can take the exterior derivative of $T^a$ and $R^a$, and use the property $dd\equiv 0$ to obtain the following two identities:
\begin{equation}
\begin{array}{lcl}
\nabla T^a & \equiv & dT^a + \omega^a{}_b \wedge T^b = R^a{}_b \wedge e^b\,, \\
\nabla R^{ab} & \equiv & dR^{ab} + \omega^a{}_c \wedge R^{cb} + \omega^b{}_c \wedge R^{ac} = 0\,. \\
\end{array}
\end{equation}
These two identities are universally valid for torsion and curvature, and are called Bianchi identities. By expanding all quantities into components as
\begin{equation}
T^a = \frac{1}{2}T^a{}_{\mu\nu} dx^{\mu} \wedge dx^{\nu} \,, \qquad
R^{ab} = \frac{1}{2}R^{ab}{}_{\mu\nu} dx^{\mu} \wedge dx^{\nu} \,,
\end{equation}
\begin{equation}
e^a = e^a{}_{\mu} dx^{\mu}\,, \qquad
\omega^{ab} = \omega^{ab}{}_{\mu} dx^{\mu}\,,
\end{equation}
and using the formula $dx^{\mu} \wedge dx^{\nu} \wedge dx^{\rho} \wedge dx^{\sigma} = \lc^{\mu\nu\rho\sigma} d^4x $, one can rewrite the Bianchi identities in component form as
\begin{equation}
\lc^{\lambda\mu\nu\rho} \left( \nabla_{\mu} T^a{}_{\nu\rho} - R^a{}_{b\mu\nu} e^b{}_{\rho} \right) = 0\,,
\end{equation}
and
\begin{equation}
\lc^{\lambda\mu\nu\rho} \nabla_{\mu} R^{ab}{}_{\nu\rho} = 0\,.
\end{equation}

For the purpose of Hamiltonian analysis, one can split the Bianchi identities into those which do not feature a time derivative and those that do. The time-independent pieces are obtained by taking $\lambda=0$ components:
\begin{equation} \label{PrviBI}
\lc^{0ijk} \left( \nabla_i T^a{}_{jk} - R^a{}_{bij} e^b{}_{k} \right) = 0\,,
\end{equation}
\begin{equation} \label{DrugiBI}
\lc^{0ijk} \nabla_i R^{ab}{}_{jk} = 0\,.
\end{equation}
These identities are valid as off-shell, strong equalities for every spacelike slice in spacetime, and can be enforced in all calculations involving the Hamiltonian analysis. The time-dependent pieces are obtained by taking $\lambda = i$ components:
\begin{equation}
\lc^{0ijk} \left( \nabla_0 T^a{}_{jk} - 2 \nabla_j T^a{}_{0k} - 2 R^a{}_{b0j} e^b{}_k - R^a{}_{bjk} e^b{}_0 \right) = 0\,,
\end{equation}
and
\begin{equation}
\lc^{0ijk} \left( \nabla_0 R^{ab}{}_{jk} - 2\nabla_j R^{ab}{}_{0k} \right) = 0\,.
\end{equation}
Due to the fact that they connect geometries of different spacelike slices in spacetime, they cannot be enforced off-shell. Instead, they can be derived from the Hamiltonian equations of motion of the theory.

In light of the Bianchi identities, we should note that the action (\ref{BFCGdejstvo}) features three more fields, $\beta^a$, $B^{ab}$ and $\phi^{ab}$, which also have field strengths $G^a$, $\nabla B^{ab}$, $\nabla \phi^{ab}$, and for which one can similarly derive Bianchi-like identities,
\begin{equation}
\nabla G^a = R^a{}_b \wedge \beta^b, \qquad \nabla^2 B^{ab} = R^a{}_c \wedge B^{cb} + R^b{}_c \wedge B^{ac}\,, \qquad \nabla^2 \phi^{ab} = R^a{}_c \wedge \phi^{cb} + R^b{}_c \wedge \phi^{ac}\,.
\end{equation}
However, due to the fact that all three fields are two-forms, in $4$-dimensional spacetime these identities will be single-component equations, with no free spacetime indices,
\begin{equation}
\lc^{\lambda\mu\nu\rho} \left( \frac{2}{3} \nabla_{\lambda} G^a{}_{\mu\nu\rho} - R^a{}_{b\mu\nu} \beta^b{}_{\nu\rho} \right) = 0\,,
\end{equation}
and similarly for $\nabla^2 B^{ab}$ and $\nabla^2 \phi^{ab}$. Therefore, these equations necessarily feature time derivatives of the fields, and do not have a purely spatial counterpart to (\ref{PrviBI}) and (\ref{DrugiBI}). In this sense, like the time-dependent pieces of the Bianchi identities, they do not enforce any restrictions in the sense of the Hamiltonian analysis, but can instead be derived from the equations of motion and expressions for the Lagrange multipliers.

\section{\label{AppC}Inverse tetrad and metric}

\hspace*{\parindent}We perform the split of the group indices into space and time components as $a = (\ubar{0},\alpha)$ where $\alpha=1,2,3$, and write the tetrad $e^a{}_{\mu}$ as a $1+3$ matrix
\begin{equation}
e^a{}_{\mu} = \left[ 
\begin{array}{c|ccc}
e^{\ubar{0}}{}_0 & & e^{\ubar{0}}{}_m & \\ \hline
 & & & \\
e^{\alpha}{}_0 & & e^{\alpha}{}_m & \\
 & & & \\
\end{array}
 \right].
\end{equation}
Then the inverse tetrad $e^{\mu}{}_b$ can be expressed in terms of the $3D$ inverse tetrad $\te{m}{\beta}$ as
\begin{equation}
e^{\mu}{}_b = \left[ 
\begin{array}{c|ccc}
\ds\frac{1}{\sigma} \vphantom{\ds\int} & &\ds - \frac{1}{\sigma} \te{m}{\beta} e^{\ubar{0}}{}_m \vphantom{\ds\int} & \\ \hline
 & & & \\
\ds -\frac{1}{\sigma} \te{m}{\alpha} e^{\alpha}{}_0 \vphantom{\ds\int} & &\ds \te{m}{\beta} + \frac{1}{\sigma} \left( \te{m}{\alpha} e^{\alpha}{}_0 \right)\left(\te{k}{\beta} e^{\ubar{0}}{}_k\right) \vphantom{\ds\int} & \\
 & & & \\
\end{array}
 \right],
\end{equation}
where
\begin{equation}
\sigma \equiv e^{\ubar{0}}{}_0 - e^{\ubar{0}}{}_k \te{k}{\alpha} e^{\alpha}{}_0
\end{equation}
is the $1\times 1$ Schur complement \cite{Zhang} of the $4\times 4$ matrix $e^a{}_{\mu}$. By definition, the $3D$ tetrad satisfies the identities
\begin{equation}
e^{\alpha}{}_m \te{m}{\beta} = \delta^{\alpha}_{\beta},\qquad
e^{\alpha}{}_m \te{n}{\alpha} = \delta^n_m.
\end{equation}
In addition, if we denote $e\equiv \det e^a{}_{\mu}$ and $\tedet\! \equiv \det e^{\alpha}{}_m$, the Schur complement $\sigma$ satisfies the Schur determinant formula
\begin{equation}
e = \sigma \tedet,
\end{equation}
which can be proved as follows.

Given any square matrix divided into blocks as
\begin{equation}
\Delta = \left[ 
\begin{array}{c|c}
A & B \\ \hline
C & M \\
\end{array}
 \right]
\end{equation}
such that $A$ and $M$ are square matrices and $M$ has an inverse, we can use the Aitken block diagonalization formula \cite{Zhang}
\begin{equation}
\left[ 
\begin{array}{c|c}
I & -BM^{-1} \\ \hline
0 & I \\
\end{array}
 \right]
\left[ 
\begin{array}{c|c}
A & B \\ \hline
C & M \\
\end{array}
 \right]
\left[ 
\begin{array}{c|c}
I & 0 \\ \hline
-M^{-1}C & I \\
\end{array}
 \right]
=
\left[ 
\begin{array}{c|c}
S & 0 \\ \hline
0 & M \\
\end{array}
 \right],
\end{equation}
where
\begin{equation}
S = A - BM^{-1}C
\end{equation}
is called the Schur complement of the matrix $\Delta$. The Aitken formula can be written in the compact form
\begin{equation}
P\Delta Q = S \oplus M,
\end{equation}
where $P$ and $Q$ are the above triangular matrices. Taking the determinant, we obtain
\begin{equation}
\det P \det\Delta \det Q = \det S \det M.
\end{equation}
Since the determinant of a triangular matrix is the product of its diagonal elements, we have $\det P = \det Q =1$, which then gives the famous Schur determinant formula:
\begin{equation}
\det\Delta = \det S \det M.
\end{equation}
Now, performing the $1+3$ block splitting of the tetrad matrix $\Delta = [e^a{}_{\mu}]_{4\times 4}$, we obtain the Schur complement $S = [\sigma ]_{1\times 1}$, while $M = [e^{\alpha}{}_m]_{3\times 3}$. The Schur determinant formula then gives
\begin{equation}
e = \sigma \tedet,
\end{equation}
which completes the proof.

Similarly to the tetrad, one can perform a $1+3$ split of the metric $g_{\mu\nu}$,
\begin{equation}
g_{\mu\nu} = \left[ 
\begin{array}{c|ccc}
g_{00} & & g_{0j} & \\ \hline
 & & & \\
g_{i0} & & g_{ij} & \\
 & & & \\
\end{array}
 \right].
\end{equation}
The inverse metric $g^{\mu\nu}$ can be expressed in terms of the $3D$ inverse metric $\tg^{ij}$ as
\begin{equation}
g^{\mu\nu} = \left[ 
\begin{array}{c|ccc}
\ds \frac{1}{\rho} \vphantom{\ds\int} & & \ds -\frac{1}{\rho} \tg^{in} g_{0i} & \\ \hline
 & & & \\
\vphantom{\ds\int} \ds -\frac{1}{\rho} \tg^{mj}g_{0j} & & \ds \tg^{mn} + \frac{1}{\rho} \left( \tg^{mj} g_{0j} \right) \left( \tg^{in} g_{0i} \right) & \\
 & & & \\
\end{array}
 \right],
\end{equation}
where
\begin{equation}
\rho \equiv g_{00} - g_{0i} \tg^{ij} g_{0j}
\end{equation}
is the $1\times 1$ Schur complement of $g_{\mu\nu}$. By definition, the $3D$ metric satisfies the identity
\begin{equation}
g_{ij} \tg^{jk} = \delta^k_i.
\end{equation}
In addition, if we denote $g\equiv \det g_{\mu\nu}$ and $\gedet \equiv \det g_{ij}$, the Schur complement $\rho$ satisfies the Schur determinant formula
\begin{equation}
g = \rho \gedet.
\end{equation}

The components of the metric can of course be written in terms of the components of the tetrad,
\begin{equation}
g_{\mu\nu} = \eta_{ab} e^a{}_{\mu} e^b{}_{\nu}.
\end{equation}
Regarding the inverse metric, the only nontrivial identity is between $\tg^{ij}$ and $\te{i}{\alpha}$. Introducing the convenient notation $\ed{\alpha} \equiv \te{i}{\alpha} e^{\ubar{0}}{}_i$, it reads:
\begin{equation}
\tg^{ij} = \te{i}{\alpha} \te{j}{\beta} \left[ \eta^{\alpha\beta} + \frac{ \eu{\alpha} \eu{\beta} }{1 - \ed{\gamma} \eu{\gamma} } \right].
\end{equation}
The relationship between determinants and Schur complements is:
\begin{equation}
g = - e^2, \qquad \gedet = \left(\!\tedet \! \right)^2 \left( 1 - \ed{\alpha} \eu{\alpha} \right) ,\qquad \rho = \frac{\sigma^2}{ \ed{\alpha} \eu{\alpha} -1}.
\end{equation}
Finally, there is one more useful identity,
\begin{equation} \label{CuveniIdentitetZaMetrikuItetradu}
g_{0j} \tg^{ij} = \te{i}{\alpha} e^{\alpha}{}_0 - \frac{\sigma}{1-\ed{\beta} \eu{\beta}} \te{i}{\alpha} \eu{\alpha}\,,
\end{equation}
which can be easily proved with some patient calculation and the other identities above.

\section{\label{AppSistemJna}Solving the system of equations}

In order to show that the constraints (\ref{NezKonstraintiProstorni}) are equivalent to the constraint (\ref{ResenaVezaZaPhi}), we proceed as follows. Introducing the shorthand notation $K_{abcd} \in \{ \lc_{abcd}, \eta_{ac} \eta_{bd} \}$, we can rewrite (\ref{NezKonstraintiProstorni}) in a convenient form
\begin{equation} \label{PolazniSistemJnaZaPhi}
e^a{}_m  K_{abcd} \lc^{0ijk} \left( e^b{}_0 R^{cd}{}_{jk}  - 2 e^b{}_j \phi^{cd}{}_{0k} \right) \approx 0\,.
\end{equation}
Next we multiply it with the Levi-Civita symbol $\lc_{0iln}$ in order to cancel the $\lc^{0ijk}$, relabel the index $m\to i$ and obtain
\begin{equation}
 K_{abcd} \left( e^a{}_i e^b{}_j \phi^{cd}{}_{0k} - e^a{}_i e^b{}_k \phi^{cd}{}_{0j}\right) \approx   K_{abcd} e^a{}_i e^b{}_0 R^{cd}{}_{jk} \,.
\end{equation}
The antisymmetrization in $jk$ indices can be eliminated by writing each equation three times with cyclic permutations of indices $ijk$, then adding the first two permutations and subtracting the third. This gives:
\begin{equation}
 K_{abcd} e^a{}_i e^b{}_j \phi^{cd}{}_{0k} \approx K_{abcd} e^a{}_0 \left[ \frac{1}{2} e^b{}_k R^{cd}{}_{ij} - e^b{}_{[i} R^{cd}{}_{j]k} \right] \,.
\end{equation}
Introducing the shorthand notation $P_{ijk}$ and $Q_{ijk}$ for the expression on the right-hand side as
\begin{equation} \label{DefinicijePiQ}
P_{ijk} \equiv \eta_{ac}\eta_{bd} e^a{}_0 \left[ \frac{1}{2} e^b{}_k R^{cd}{}_{ij} - e^b{}_{[i} R^{cd}{}_{j]k} \right]\,, \qquad
Q_{ijk} \equiv \lc_{abcd} e^a{}_0 \left[ \frac{1}{2} e^b{}_k R^{cd}{}_{ij} - e^b{}_{[i} R^{cd}{}_{j]k} \right]\,,
\end{equation}
our system can be rewritten as
\begin{equation} \label{SistemPrekoPiQ}
\eta_{ac}\eta_{bd} e^a{}_i e^b{}_j \phi^{cd}{}_{0k} \approx P_{ijk}\,, \qquad
\lc_{abcd} e^a{}_i e^b{}_j \phi^{cd}{}_{0k} \approx Q_{ijk}\,.
\end{equation}
This system consists of $18$ equations for the $18$ variables $\phi^{ab}{}_{0k}$. We look for a solution in the form
\begin{equation} \label{OblikResenjaZaPhi}
\phi^{cd}{}_{0k} = A^{cdmn} P_{mnk} + B^{cdmn} Q_{mnk}\,,
\end{equation}
where the coefficients $A^{cdmn}$ and $B^{cdmn}$ are to be determined, for arbitrarily given values of $P_{ijk}$ and $Q_{ijk}$. Substituting (\ref{OblikResenjaZaPhi}) into (\ref{SistemPrekoPiQ}) we obtain
\begin{equation}
\begin{array}{lcl}
\Big[ \eta_{ac}\eta_{bd} e^a{}_i e^b{}_j A^{cdmn} - \delta^{[m}_i \delta^{n]}_j \Big] P_{mnk} + \Big[ \eta_{ac}\eta_{bd} e^a{}_i e^b{}_j B^{cdmn} \Big] Q_{mnk} & \approx & 0\,, \\
\Big[ \lc_{abcd} e^a{}_i e^b{}_j A^{cdmn} \Big] P_{mnk} + \Big[ \lc_{abcd} e^a{}_i e^b{}_j B^{cdmn} - \delta^{[m}_i \delta^{n]}_j  \Big] Q_{mnk} & \approx & 0\,. \\
\end{array}
\end{equation}
Since $P_{mnk}$ and $Q_{mnk}$ are considered arbitrary, the expressions in the brackets must vanish, giving the following equations for $A^{cdmn}$,
\begin{equation} \label{SistemJnaZaAkoef}
\eta_{ac}\eta_{bd} e^a{}_i e^b{}_j A^{cdmn} \approx \delta^{[m}_i \delta^{n]}_j \,, \qquad \lc_{abcd} e^a{}_i e^b{}_j A^{cdmn} \approx 0\,,
\end{equation}
and for $B^{cdmn}$,
\begin{equation} \label{SistemJnaZaBkoef}
 \eta_{ac}\eta_{bd} e^a{}_i e^b{}_j B^{cdmn} \approx 0 \,, \qquad
\lc_{abcd} e^a{}_i e^b{}_j B^{cdmn} \approx \delta^{[m}_i \delta^{n]}_j \,.
\end{equation}

Focus first on (\ref{SistemJnaZaAkoef}). The first equation can be rewritten in the form
\begin{equation} \label{PrvaJednacinaZaAkoef}
e_{ci} e_{dj} A^{cdmn} \approx \delta^{[m}_i \delta^{n]}_j \,,
\end{equation}
and we want to rewrite the second equation in a similar form as well. In order to do that, we need to get rid of the Levi-Civita symbol on the left-hand side, by virtue of the identity
\begin{equation}
\det(e_{a\mu}) \lc_{abcd} = \lc^{\mu\nu\rho\sigma} e_{a\mu} e_{b\nu} e_{c\rho} e_{d\sigma} \,.
\end{equation}
Noting that $\det (e_{a\mu}) = \det (\eta_{ab}e^b{}_{\mu}) = -\det(e^a{}_{\mu}) = -e$ and introducing the metric $g_{\mu\nu} \equiv e^a{}_{\mu} e_{a\nu}$, we can multiply this identity with $e^a{}_i e^b{}_j$ to obtain:
\begin{equation}
\lc_{abcd} e^a{}_i e^b{}_j = -\frac{1}{e} \lc^{\mu\nu\rho\sigma} g_{\mu i} g_{\nu j} e_{c\rho} e_{d\sigma}\,.
\end{equation}
Substituting this into the second equation in (\ref{SistemJnaZaAkoef}) gives
\begin{equation}
\lc^{\mu\nu\rho\sigma} g_{\mu i} g_{\nu j} e_{c\rho} e_{d\sigma} A^{cdmn} \approx 0\,.
\end{equation}
Next we expand the $\rho$ and $\sigma$ indices into space and time components as $\rho=(0,k)$ and $\sigma=(0,l)$ to obtain
\begin{equation}
2\lc^{\mu\nu 0l} g_{\mu i} g_{\nu j} e_{c0} e_{dl} A^{cdmn} +
\lc^{\mu\nu kl} g_{\mu i} g_{\nu j} e_{ck} e_{dl} A^{cdmn} \approx 0\,.
\end{equation}
The second term on the left can be evaluated using (\ref{PrvaJednacinaZaAkoef}), which gives:
\begin{equation}
2\lc^{\mu\nu 0l} g_{\mu i} g_{\nu j} e_{c0} e_{dl} A^{cdmn} +
\lc^{\mu\nu mn} g_{\mu i} g_{\nu j} \approx 0\,.
\end{equation}
The Levi-Civita symbol in the first term is nonzero only if $\mu\nu$ are spatial indices, so we can write
\begin{equation} \label{JnaSesnaest}
2\lc^{rs0l} g_{ri} g_{sj} e_{c0} e_{dl} A^{cdmn} +
\lc^{\mu\nu mn} g_{\mu i} g_{\nu j} \approx 0\,.
\end{equation}
At this point we need to introduce $3D$ inverse metric, $\tg^{ij}$, and to split the group indices into $3+1$ form $a=(\ubar{0},\alpha)$ where $\alpha=1,2,3$, see Appendix \ref{AppC}. Multiplying (\ref{JnaSesnaest}) with two inverse spatial metrics and another Levi-Civita symbol, we can finally rewrite it as:
\begin{equation}
e_{c0} e_{di} A^{cdmn} \approx g_{0j} \tg^{j[m} \delta^{n]}_i \,.
\end{equation}
The goal of all these transformations was to rewrite the system (\ref{SistemJnaZaAkoef}) into the form
\begin{equation}
e_{ci} e_{dj} A^{cdmn} \approx \delta^{[m}_i \delta^{n]}_j \,, \qquad
e_{c0} e_{di} A^{cdmn} \approx g_{0j} \tg^{j[m} \delta^{n]}_i \,.
\end{equation}
At this point we can expand the group indices on the left-hand side into $3+1$ form, to obtain:
\begin{equation} \label{JnaDevetnaest}
e_{\gamma i} e_{\delta j} A^{\gamma\delta mn} + \left( e^{\ubar{0}}{}_j e_{\delta i} - e^{\ubar{0}}{}_i e_{\delta j} \right) A^{\ubar{0}\delta mn} \approx \delta^{[m}_i \delta^{n]}_j \,,
\end{equation}
\begin{equation} \label{JnaDvadeset}
e_{\gamma 0} e_{\delta j} A^{\gamma\delta mn} + \left( e^{\ubar{0}}{}_j e_{\delta 0} - e^{\ubar{0}}{}_0 e_{\delta j} \right) A^{\ubar{0}\delta mn} \approx g_{0k} \tg^{k[m} \delta^{n]}_j \,.
\end{equation}
Now we multiply (\ref{JnaDevetnaest}) with $\te{i}{\alpha} e^{\alpha}{}_0$ and subtract it from (\ref{JnaDvadeset}). The first terms on the left cancel, and (\ref{JnaDvadeset}) becomes
\begin{equation}
-\sigma e_{\delta j} A^{\ubar{0}\delta mn} \approx g_{0k} \tg^{k[m} \delta^{n]}_j - \te{[m}{\alpha} \delta^{n]}_j e^{\alpha}{}_0 \,,
\end{equation}
where $\sigma$ is the $1\times 1$ Schur complement matrix of the tetrad $e^a{}_{\mu}$ (see Appendix \ref{AppC}). Multiplying with another inverse $3D$ tetrad and using the identity (\ref{CuveniIdentitetZaMetrikuItetradu}), we finally obtain the first half of the coefficients $A$:
\begin{equation} \label{RnjeZaAkoefPrvaPolovina}
A^{\ubar{0}\alpha mn} \approx \frac{1}{1- \eu{\gamma} \ed{\gamma}} \te{[m}{\delta} \teu{n]}{\alpha} \eu{\delta}\,.
\end{equation}
Finally, substituting this back into (\ref{JnaDevetnaest}) and multiplying with two more inverse $3D$ tetrads we obtain the second half of the coefficients $A$:
\begin{equation} \label{RnjeZaAkoefDrugaPolovina}
A^{\alpha\beta mn} \approx \teu{[m}{\alpha} \teu{n]}{\beta} + \frac{\eu{\delta}}{1-\ed{\gamma}\eu{\gamma}} \left[ \eu{\alpha} \te{[m}{\delta} \teu{n]}{\beta} - \eu{\beta} \te{[m}{\delta} \teu{n]}{\alpha} \right] \,.
\end{equation}

Next we turn to the system (\ref{SistemJnaZaBkoef}) for coefficients $B$. The method to solve it is completely analogous to the above method of solving (\ref{SistemJnaZaAkoef}), and we will not repeat all the steps, but rather only quote the final result:
\begin{equation} \label{RnjeZaBkoefPrvaPolovina}
B^{\ubar{0}\beta mn} \approx \frac{1}{4} \lc^{\ubar{0}\beta\gamma\delta} \left[ \te{m}{\gamma} \te{n}{\delta} + 2 \te{[m}{\alpha} \te{n]}{\delta} \frac{\eu{\alpha} \ed{\gamma}}{1-\ed{\epsilon}\eu{\epsilon}} \right] \,,
\end{equation}
and
\begin{equation} \label{RnjeZaBkoefDrugaPolovina}
B^{\alpha\beta mn} \approx \frac{1}{2} \, \frac{1}{1-\ed{\epsilon}\eu{\epsilon}} \, \lc^{\ubar{0}\alpha\beta\gamma} \te{[m}{\gamma} \te{n]}{\delta} \eu{\delta} \,.
\end{equation}

To conclude, by determining the $A$ and $B$ coefficients in (\ref{OblikResenjaZaPhi}) we have managed to solve the original system of equations (\ref{PolazniSistemJnaZaPhi}) for $\phi^{ab}{}_{0k}$. Substituting (\ref{DefinicijePiQ}) into (\ref{OblikResenjaZaPhi}) the expression for $\phi^{ab}{}_{0k}$ can be arranged into the form
\begin{equation} \label{ResenjeZaPhiUzgodnomObliku}
\phi^{ab}{}_{0k} \approx e^f{}_0 R^{cd}{}_{mn} F^{abmn}{}_{fcdk}\,,
\end{equation}
where
\begin{equation} \label{KonacnaJnaZaVelikoF}
F^{abmn}{}_{fcdk} \equiv \frac{1}{2} \Big[ A^{abmn} \eta_{fc} e_{dk} -2 A^{abim} \eta_{fc} e_{di} \delta^n_k + B^{abmn} \lc_{fhcd} e^h{}_k - 2 B^{abim} \lc_{fhcd} e^h{}_i \delta^n_k \Big] \,,
\end{equation}
and coefficients $A$ and $B$ are specified by (\ref{RnjeZaAkoefPrvaPolovina}), (\ref{RnjeZaAkoefDrugaPolovina}), (\ref{RnjeZaBkoefPrvaPolovina}) and (\ref{RnjeZaBkoefDrugaPolovina}). Note that (\ref{KonacnaJnaZaVelikoF}) depends only on $e^a{}_i$ components of the metric (in a very complicated way), while the dependence of $\phi^{ab}{}_{0k}$ on $e^a{}_0$ and $\omega^{ab}{}_i$ is factored out in (\ref{ResenjeZaPhiUzgodnomObliku}).

\section{\label{AppLeviCivitaIdentity}Levi-Civita identity}

The identity for the Levi-Civita symbol in $4$ dimensions used in the main text is:
\begin{equation} \label{LeviCivitaIdentity}
A_{[a} \lc_{b]cdf} C^c D^d F^f = - \frac{1}{2} \lc_{abcd} A_f \left[ C^d D^f F^c + C^c D^d F^f + C^f D^c F^d \right].
\end{equation}
The proof goes as follows. Denote the left-hand side of the identity as
\begin{equation}
K_{ab} \equiv A_{[a} \lc_{b]cdf} C^c D^d F^f
\end{equation}
and take the dual to obtain:
\begin{equation}
\lc^{aba'b'} K_{ab} = \lc^{aba'b'} \lc_{bcdf} A_a C^c D^d F^f.
\end{equation}
Next expand the product of two Levi-Civita symbols into Kronecker deltas and use them to contract the vectors $A$, $C$, $D$ and $F$:
\begin{equation}
\lc^{aba'b'} K_{ab} = 2\left[ (A\cdot D) F^{[a'} C^{b']} + (A\cdot F) C^{[a'} D^{b']} + (A\cdot C) D^{[a'} F^{b']} \right] .
\end{equation}
Now take the dual again, i.e. contract with $\lc_{a'b'cd}$ to obtain
\begin{equation}
-4 K_{cd} = \lc_{a'b'cd} \lc^{aba'b'} K_{ab} = 2\lc_{a'b'cd} \left[ (A\cdot D) F^{[a'} C^{b']} + (A\cdot F) C^{[a'} D^{b']} + (A\cdot C) D^{[a'} F^{b']} \right] .
\end{equation}
Finally, multiply by $-1/4$ and relabel the indices to obtain
\begin{equation}
K_{ab} = - \frac{1}{2} \lc_{abcd} A_f \left[ C^d D^f F^c + C^c D^d F^f + C^f D^c F^d \right],
\end{equation}
which proves the identity.

\section{\label{AppE}Relation between the $BFCG$ and the MacDowell-Mansouri models}

Given that the constrained $BFCG$ action (\ref{BFCGdejstvo}) is equivalent to GR, it is a straightforward exercise to include a cosmological constant term:
\begin{equation} \label{BFCGdejstvoSaCC}
S_{GR\Lambda} = \int_{\cM} \left[B_{ab} \wedge R^{ab} + e^a \wedge G_a - \phi^{ab} \wedge \left( B_{ab} - \lc_{abcd}\, e^c\wedge e^d \right) - \frac{\Lambda}{6} \, \lc_{abcd} \, e^a \wedge e^b \wedge e^c \wedge e^d  \right]\,,
\end{equation}
Working out the corresponding equations of motion, one obtains the same set (\ref{JKzaPhiBolja}), (\ref{JKzaBbolja}), (\ref{JKzaBetaBolja}), (\ref{JKzaTbolja}) as for the action (\ref{BFCGdejstvo}), except for the Einstein field equation (\ref{JKzaEiRbolja}) which is modified into
\begin{equation}
\lc_{abcd} \left( R^{bc} - \frac{\Lambda}{3}\, e^b \wedge e^c \right) \wedge e^d = 0\,,
\end{equation}
which can in turn be rewritten into the standard component form
\begin{equation}
R_{\mu\nu} - \frac{1}{2} R\, g_{\mu\nu} + \Lambda \, g_{\mu\nu} = 0\,.
\end{equation}
Here the parameter $\Lambda\in\realni$ is the cosmological constant.

It is interesting to note that one can obtain the MacDowell-Mansouri action for GR \cite{MacDowellMansouri,Smolin,LingSmolin,SmolinStarodubtsev,Wise} from the action (\ref{BFCGdejstvoSaCC}). In particular, the relationship between (\ref{BFCGdejstvoSaCC}) and the MacDowell-Mansouri action is analogous to the relationship between the Palatini and Einstein-Hilbert actions, respectively, as we shall now demonstrate. To this end, first add and subtract a term $\zeta B_{ab}\wedge e^a \wedge e^b$ to (\ref{BFCGdejstvoSaCC}), where $\zeta = \pm 1$, and rewrite it in the form
\begin{equation} 
S_{GR\Lambda} = \int_{\cM} \left[B_{ab} \wedge \left( R^{ab} - \zeta e_a \wedge e_b \right) + e^a \wedge G_a - \phi^{ab} \wedge \left( B_{ab} - \lc_{abcd}\, e^c\wedge e^d \right) + e^a \wedge e^b \wedge \left( \zeta B_{ab} - \frac{\Lambda}{6} \lc_{abcd}\, e^c \wedge e^d  \right) \right].
\end{equation}
Next we perform the partial integration over the $e^a \wedge G_a$ term, and rewrite the action as
\begin{equation} \label{MDMactionHalfway}
S_{GR\Lambda} = \int_{\cM} \left[B_{ab} \wedge \left( R^{ab} - \zeta e_a \wedge e_b \right) + \beta^a \wedge \nabla e_a - \phi^{ab} \wedge \left( B_{ab} - \lc_{abcd}\, e^c\wedge e^d \right) + e^a \wedge e^b \wedge \left( \zeta B_{ab} - \frac{\Lambda}{6} \lc_{abcd}\, e^c \wedge e^d  \right) \right].
\end{equation}
Now we want to eliminate the Lagrange multiplier $\phi^{ab}$ from the action. This is performed in analogy with the way the Palatini action is transformed into the Einstein-Hilbert action --- we take the variation of the action with respect to $\phi^{ab}$ to obtain the corresponding equation of motion, and then substitute this equation back into the action. The equation of motion is algebraic rather than differential,
\begin{equation}
B_{ab} = \lc_{abcd} e^c \wedge e^d\,,
\end{equation}
which suggests that no propagating degrees of freedom will be lost upon substituting it back into the action. So we solve it for the product of two tetrads,
\begin{equation}
e^a \wedge e^b = -\frac{1}{4} \lc^{abcd} B_{cd}\,,
\end{equation}
and substitute it back into (\ref{MDMactionHalfway}), eliminating the product of the tetrads from all terms except the first one, to obtain:
\begin{equation} \label{MDMactionNoncovariant}
S = \int_{\cM} \left[B_{ab} \wedge \left( R^{ab} - \zeta e_a \wedge e_b \right) + \beta^a \wedge \nabla e_a + \frac{\Lambda - 6\zeta}{24} \lc_{abcd}\, B^{ab} \wedge B^{cd} \right]\,.
\end{equation}
Note that the term containing $\phi^{ab}$ has vanished from the action, while the final term has been transformed into the $B\wedge B$ term.

Finally, to see that (\ref{MDMactionNoncovariant}) is actually the MacDowell-Mansouri action, introduce the following change of variables:
\begin{equation}
B^{AB} \equiv \left[
\begin{array}{ccc|c}
 & & & \\
 & B^{ab} & & \frac{\beta^a}{2} \\
 & & & \\ \hline
 & -\frac{\beta^b}{2} & & 0 \\
\end{array}
\right]\,, \qquad A^{AB} \equiv \left[
\begin{array}{ccc|c}
 & & & \\
 & \omega^{ab} & & e^a \\
 & & & \\ \hline
 & -e^b & & 0 \\
\end{array}
\right]\,, 
\end{equation}
and
\begin{equation}
F^{AB} \equiv dA^{AB} + A^A{}_C \wedge A^{CB} = \left[ 
\begin{array}{ccc|c}
 & & & \\
 & R^{ab} - \zeta e^a \wedge e^b & & \nabla e^a \\
 & & & \\ \hline
 & - \nabla e^b & & 0 \\
\end{array}
\right]\,, \qquad V^A \equiv \left[
\begin{array}{c}
 0 \\
 0 \\
 0 \\
 0 \\ \hline
 1 \\
\end{array}
 \right]\,.
\end{equation}
These represent the $5$-dimensional $2$-form $B^{AB}$, connection $1$-form $A^A$, its field strength $2$-form $F^{AB}$ and a $0$-form $V^A$. The capital Latin indices take values $0,1,2,3,5$, and we can also introduce the $5$-dimensional Levi-Civita symbol $\lc_{ABCDE}$, which is related to the ordinary $4$-dimensional one as $\lc_{abcd5} \equiv \lc_{abcd}$. Using all this, the action (\ref{MDMactionNoncovariant}) can be rewritten into the form
\begin{equation} \label{MDMactionCovariant}
S = \int_{\cM} \left[B_{AB} \wedge F^{AB} + \frac{\Lambda - 6\zeta}{24} \, B^{ab} \wedge B^{cd} \, \lc_{ABCDE} \, V^E \right]\,,
\end{equation}
which is manifestly covariant with respect to the action of the groups $SO(4,1)$ or $SO(3,2)$, depending on the choice of $\zeta = \pm 1$, which enters the $5$-dimensional metric
\begin{equation}
\eta_{AB} \equiv \left[ 
\begin{array}{cccc|c}
 -1 &   &   &   &       \\
    & 1 &   &   &       \\
    &   & 1 &   &       \\
    &   &   & 1 &       \\ \hline
    &   &   &   & \zeta \\
\end{array}
 \right]\,,
\end{equation}
where the off-diagonal values are assumed to be zero. The action (\ref{MDMactionCovariant}) is precisely the $BF$-formulation of the MacDowell-Mansouri action \cite{MacDowellMansouri,Smolin,LingSmolin,SmolinStarodubtsev,Wise}, as we have set out to demonstrate.

\end{document}